\documentclass[conference]{IEEEtran}
\IEEEoverridecommandlockouts
\usepackage{cite}
\usepackage{amsmath,amssymb,amsfonts}
\usepackage{graphicx}
\usepackage{textcomp}
\usepackage{xcolor}
\usepackage{caption}
\usepackage{subcaption}
\usepackage{xspace}

\usepackage{algorithm}
\usepackage[noend]{algorithmic}

\newcommand{\sol}{\textsc{CiFold}\xspace}

\begin{document}
 
\title{Circuit Folding: Scalable and Graph-Based Circuit Cutting via Modular Structure Exploitation
}
\vspace{-1em}
\author{
Shuwen Kan$^{1}$,
Yanni Li$^{1}$,
Hao Wang$^{2}$,
Sara Mouradian$^{3}$,
Ying Mao$^{1}$\\
$^{1}$Fordham University, New York, NY, USA\\
$^{2}$Stevens Institute of Technology, Hoboken, NJ, USA\\
$^{3}$University of Washington, Seattle, WA, USA\\
\{sk107, yl32, ymao41\}@fordham.edu, hwang9@stevens.edu, smouradi@uw.edu
}

\IEEEaftertitletext{\vspace{-2em}}

\maketitle
\begin{abstract}
Circuit cutting is a promising technique that leverages both quantum and classical computational resources, enabling the practical execution of large quantum circuits on noisy intermediate-scale quantum (NISQ) hardware. Recent approaches typically focus exclusively on either gate cuts or wire cuts, modeling quantum circuits as graphs. However, identifying optimal cutting locations using this representation often results in prohibitively high computational complexity, especially under realistic hardware constraints.
In this paper, we introduce \sol, a novel graph-based framework that exploits repetitive modular structures inherent in quantum algorithms, significantly enhancing the scalability and efficiency of circuit cutting. Our approach systematically folds quantum circuits into compact meta-graphs by identifying and merging common gate sequences across entangled qubits, dramatically simplifying subsequent partitioning tasks. We define folding factor and  variance to quantify circuit compression and ensure balanced folding. Using these condensed representations, \sol precisely identifies cut locations without exhaustive global graph searches.
We perform extensive experiments, comparing \sol with state-of-the-art circuit-cutting techniques. Results demonstrate that \sol achieves superior partition quality and computational efficiency, reducing the number of required cuts by an average of $31.6\%$ and lowering the sampling overhead substantially by $3.55 \times 10^{9}$. Our findings illustrate that \sol represents a significant advancement toward scalable quantum circuit cutting.

 \end{abstract}
\vspace{-0.5em}

\section{Introduction}

Quantum computing holds immense potential by solving complex problems currently intractable for classical computers. However, current noisy intermediate-scale quantum (NISQ) devices suffer from limited qubit counts, connectivity constraints, and gate fidelity issues, severely hindering the practical execution of large-scale quantum circuits.

Circuit cutting has recently emerged as a promising strategy to extend quantum computation beyond current hardware limitations by decomposing large circuits into smaller subcircuits executable on resource-constrained quantum processors. Two main circuit-cutting paradigms: wire cut~\cite{tang2025tensorqcscalabledistributedquantum,Tang_2021,lowe2023fast}, which provides exact reconstruction but incurring fixed overhead, and gate cut~\cite{ren2024hardware,ufrecht2023cutting,schmitt2024cutting}, which uses quasiprobability decomposition to reconstruct expectation values with reduced sampling overhead compared to wire cut.

Despite the complementary nature of gate- and wire-cutting methods, their integration into a unified framework remains largely unexplored. Recent studies have begun examining the advantages of combining these two approaches~\cite{brandhofer2023optimal,pawar2024qrcc}, but finding optimal cut points within a hybrid gate-wire partitioning is inherently challenging. Existing graph-based techniques for wire cutting represent each two-qubit gate as a node~\cite{Tang_2021,kan2024scalable}, whereas gate cutting adopts a simpler representation, using each qubit as a node~\cite{ren2024hardware}. When integrating both gate and wire cuts into one framework, the resulting graph structure effectively requires twice as many nodes compared to the wire-cut-only scenario. Consequently, solver-based methods designed to identify optimal partitions (e.g., Satisfiability Modulo Theories(SMT) solvers used in~\cite{brandhofer2023optimal} or Integer Linear Programming(ILP) solvers employed in~\cite{pawar2024qrcc}) face significant runtime overhead. For instance, Brandhofer et al.\cite{brandhofer2023optimal} imposed a one-hour SMT solver timeout for circuits of up to 40 qubits, while Pawar et al.\cite{pawar2024qrcc}, using an improved ILP model, reported solution times of up to 1800 seconds for a 20-qubit quantum Fourier transform (QFT) circuit. They suffer from significant scalability issues and are impractical in reality. 

Motivated by the fact that many quantum algorithms follow fixed structural patterns to solve well-defined problems such as Grover’s algorithm \cite{grover1996fast}, quantum phase estimation\cite{nielsen2010quantum}, quantum arithmetic operation\cite{cuccaro2004new}, Bernstein-Vazirani algorithm\cite{bernstein1993quantum} and Shor's algorithm\cite{shor1994algorithms}. As problem size increases, these algorithms typically scale by adding more qubits while retaining the same circuit structure. While this modularity offers an opportunity for optimization, current approaches often treat each gate as unique, leading to a combinatorial explosion in cut-point searches that is both non-optimal and computationally expensive. This results in non-optimal or time-consuming methods that fail to scale efficiently. 
To address these limitations, this work proposes \emph{\sol}, a novel folding-based circuit-cutting framework designed to harness the inherent modularity of many quantum algorithms. Rather than searching the entire circuit for cutting points, \sol identifies recurring gate sequences and exploits these structural redundancies to minimize the number of required cuts. By incorporating self-adaptive algorithms tailored to detect, from qubit-level, in parallel, and leverage repeated modules, \sol significantly reduces the exponential overhead commonly associated with circuit partitioning. Experimental results on standard quantum benchmarks demonstrate how our approach not only lowers the computational cost but also preserves accuracy, offering a practical path forward for large-scale quantum computations. 
This work introduces several key innovations:
\begin{itemize}
    \item \textbf{Graph-Based Circuit Folding:} We present a novel graph-based framework for circuit partitioning that models quantum circuits as directed graphs. By applying folding techniques to identify recurring gate sets, \sol~constructs a meta-graph that reduces circuit complexity and improves partitioning efficiency.
    
    \item \textbf{Folding Metrics:} We define the folding factor and folding variance as quantitative measures of circuit folding efficiency, providing a systematic evaluation of the gains achieved through structured circuit decomposition.
    
    \item \textbf{Adaptive Partitioning:} \sol leverages the meta-graph structure to guide the search for optimal partition points, efficiently reducing the number of subcircuit executions while maintaining fidelity. This adaptive approach ensures flexibility across varying hardware constraints and qubit resources.
    
    \item \textbf{Implementation and Evaluation:} We implement end-to-end circuit cutting pipeline using \sol  and evaluate it across a diverse set of quantum circuit benchmarks. \sol improves relative fidelity, ranging from 5.3\% to 61.2\% across seven IBM backend emulators, achieves runtimes below 1-second even for complex circuits,and reduces required cuts by an average of 31.6\%, leading to a sampling overhead reduction of \(3.55 \times 10^{9}\).  

\end{itemize}

\section{Related Work and Background }
\label{sec:background}

Quantum circuit cutting broadly falls into two categories: theoretical explorations, primarily investigating bipartite cuts~\cite{mitarai2021constructing,Mitarai2021overheadsimulating,brenner2023optimal,piveteau2023circuit,lowe2023fast,PRXQuantum.5.040308,pednault2023alternative,chen2022approximate,chen2023efficient}, and practical, implementation-driven studies aimed at efficient real-world deployment~\cite{Tang_2021,tang2025tensorqcscalabledistributedquantum,kan2024scalable,ren2024hardware,Gentinetta2024overheadconstrained,li2024casequantumcircuitcutting,qiskit-addon-cutting,li2024qutracermitigatingquantumgate}. Early theoretical work employed quasiprobability decomposition, analyzing overhead scaling and techniques such as parallel cuts and classical communication. Practical implementations like CutQC~\cite{Tang_2021}, TensorQC~\cite{tang2025tensorqcscalabledistributedquantum}, and FitCut~\cite{kan2024scalable} introduced automated pipelines addressing classical computation bottlenecks and hardware-aware optimization. 

In contrast to prior work, \sol~framework uniquely exploits qubit-level structural repetition within quantum circuits, using a graph-based folding method to generate compact meta-graphs. This approach reduces computational overhead, enables parallel subcircuit processing, and efficiently optimizes cut selection for both quantum and classical resources.

\subsection{Circuit Cutting}
\begin{figure}
    \centering
    \includegraphics[width=1\linewidth]{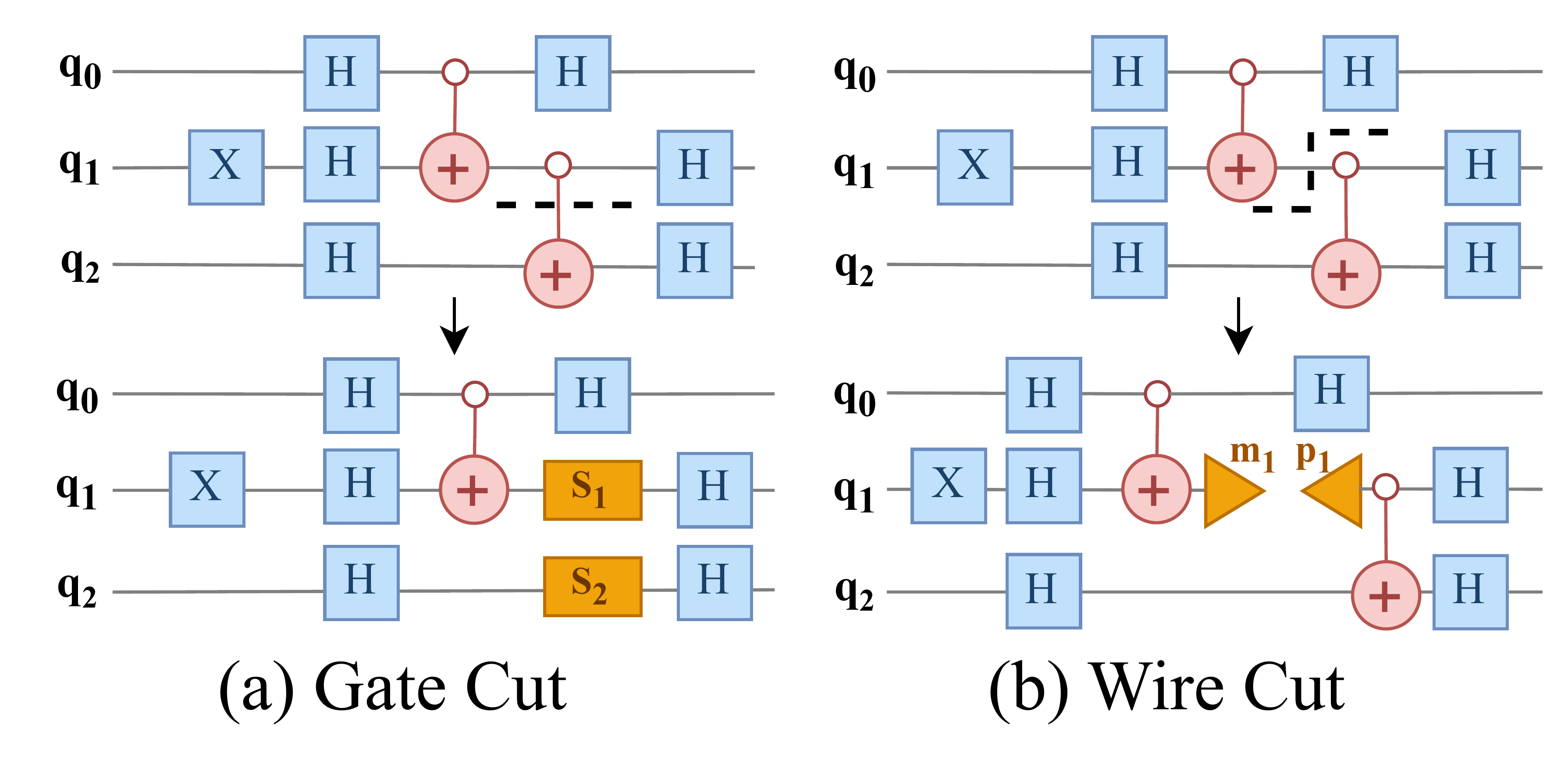}
    \caption{Illustration of Gate Cut and Wire Cut}
    \label{fig:gate_wire}
    \vspace{-0.2in}
\end{figure}

Quantum circuit cutting allows large quantum circuits to be divided into smaller subcircuits that can be executed independently on quantum hardware and later reconstructed using classical postprocessing. Two main techniques have been developed for this purpose: gate cutting and wire cutting, as shown in Figure~\ref{fig:gate_wire}. 


\textbf{Gate cut}: employs quasiprobability simulation to reconstruct the expectation value of the original circuit. It replaces a two-qubit gate with local single-qubit operations, denoted by $s_i$ as shown in Fig~\ref{fig:gate_wire}, along with classical postprocessing~\cite{mitarai2021constructing}. Each gate cut results in separately executable subcircuits, which are individually sampled according to their corresponding quasiprobability distribution.
\textbf{Wire cuts}: A wire cut severs a quantum wire by inserting measurement and preparation operations that effectively separate a circuit into independently executable fragments. As shown in Fig.~\ref{fig:gate_wire}(b), when two wires are cut, $m_1$ and $m_2$ correspond to measurement channels in the $\{I, X, Y, Z\}$ bases, while $p_1$ and $p_2$ represent state preparations in the $\{\lvert 0 \rangle, \lvert 1 \rangle, \lvert i \rangle, \lvert + \rangle\}$ states for a distinct computation fragment~\cite{Tang_2021}. Consequently, one cut entails executing three different upstream subcircuits (since $I$ and $Z$ are identical) and four different downstream subcircuits. The final outputs are combined via 16 Kronecker products of the respective probability distributions to achieve exact reconstruction.

\vspace{-.5\baselineskip}

\subsection{End-to-End Circuit Cutting pipeline}
\label{subsec:background_piepline}

Circuit cutting is a hybrid quantum-classical technique that redistributes computational workload between quantum hardware and classical postprocessing. To enable its practical deployment, an end-to-end workflow is required that coordinates quantum execution with classical preprocessing and reconstruction. This workflow consists of three key stages:

\textbf{1) Cut Point Identification:}
The first step is to determine the cut locations within the quantum circuit by jointly considering hardware constraints and the requirements of circuit cutting protocols. While the maximum number of qubits per subcircuit is typically the primary constraint, other hardware limitations—such as gate depth restrictions imposed by coherence time and gate fidelity—must also be included in the cut selection process.

The cost model must reflect the specific circuit cutting protocol in use. For example, gate cuts and wire cuts differ significantly in their reconstruction overhead and operational requirements. Techniques like parallel cutting can reduce joint sampling overhead between subcircuits \cite{brenner2023optimal, schmitt2024cutting}, but they introduce additional constraints and trade-offs that must be explicitly accounted for in the model.
The overall objective is to minimize the total sampling overhead required for reconstruction, which can scale exponentially with the number and type of cuts. Therefore, identifying an optimal cut strategy requires careful alignment between hardware feasibility and protocol-specific cost modeling. 

\textbf{2) Subcircuit Execution:}
The second step involves compiling and executing subcircuits. Wire cutting uses a fixed number of shots per subcircuit, while gate cutting requires uneven shot allocation based on quasiprobability weights to ensure statistical accuracy.

\textbf{3) Classical Postprocessing:}
Following quantum execution, classical reconstruction is performed to assemble either the full probability distribution (in the case of wire cutting) or the expectation value of specified observables (in the case of gate cutting). This reconstruction step follows the protocol determined during the initial cut planning phase.

\section{Challenges and Opportunities} \label{subsec:challenge}
\begin{figure}
   \centering
   \includegraphics[width=1\linewidth]{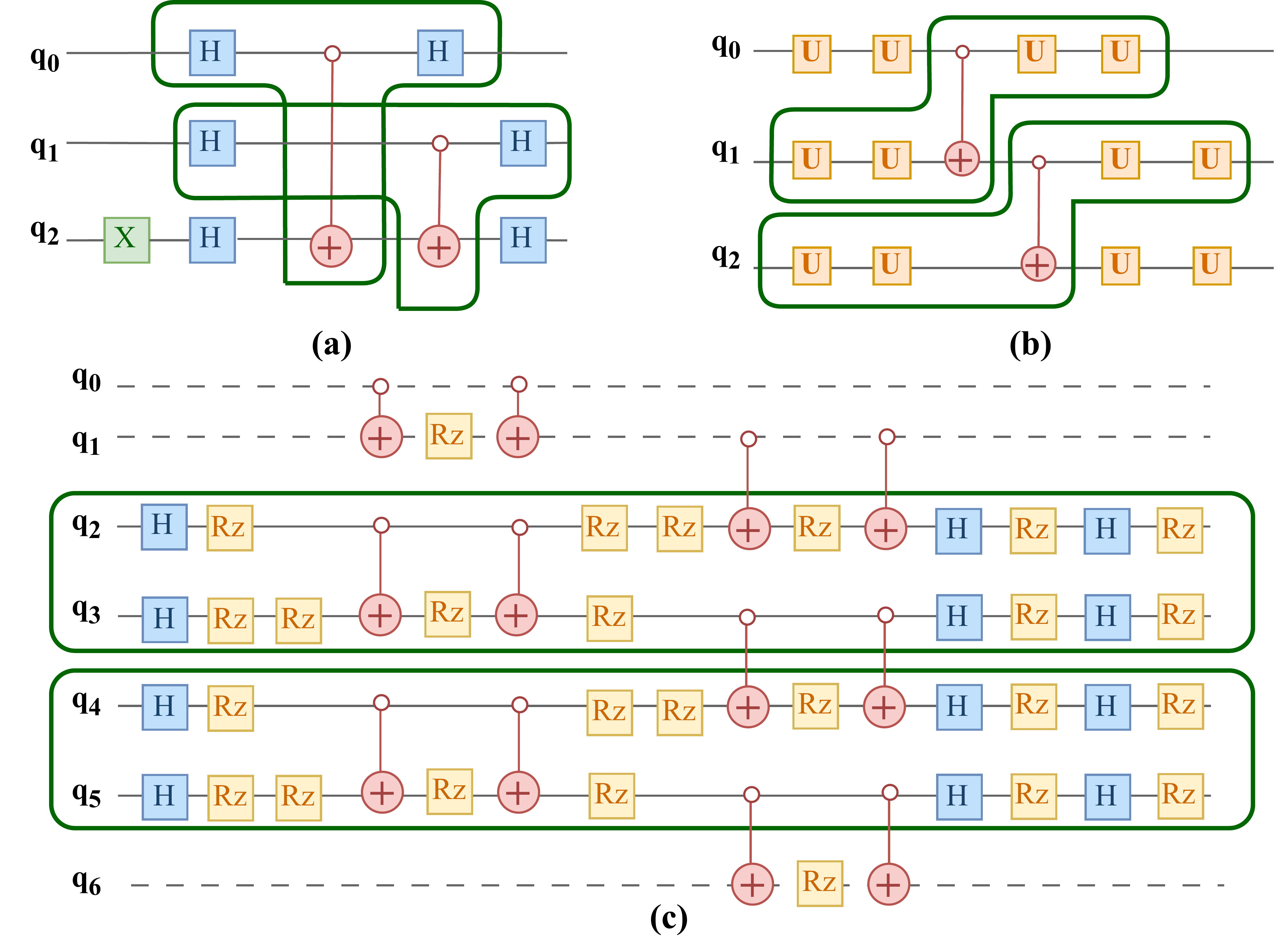}
   \caption{Circuits with repeated patterns: (a) Bernstein Vazirani (BV); (b) Hardware efficient ansatz (HWEA); (c) ISING}
   \vspace{-0.15in}
   \label{fig:motivation}
\end{figure}

\noindent\textbf{Insight 1: Cut Finding is Non Trivial --- }
Despite its critical role, the problem of cut point finding is often overlooked or assumed to be predefined in many circuit cutting studies\cite{schmitt2024cutting, lowe2023fast, ufrecht2023cutting, ufrecht2024optimal}. Theoretical works frequently operate under the assumption that optimal cuts are known a priori. In practice, determining where to cut is a complex optimization task that cannot be easily addressed using well established graph algorithms. To address this, existing implementations typically formulate cut finding as a constrained optimization problem, employing solvers such as Mixed Integer Programming (MIP), Boolean Satisfiability (SAT), or Integer Linear Programming (ILP) to model sampling overhead and search for globally optimal partitions \cite{Tang_2021, pawar2024qrcc, brandhofer2023optimal}.

The complexity of the graph model differs greatly between gate and wire cutting. Gate cutting's graph representations map nodes to qubits, enabling heuristic-based merging strategies~\cite{ren2024hardware}. In contrast, wire cuts model nodes as two-qubit gates, making valid partitioning under qubit constraints a factorial-complexity problem~\cite{Tang_2021, kan2024scalable}. Hybrid gate-wire modeling further doubles graph size and increases solver complexity. Even with optimized ILP formulations \cite{pawar2024qrcc}, solving times on practical circuits can exceed 1800 seconds, underscoring the need for more scalable and hardware-aware heuristics.

These challenges present a key opportunity for future research: developing efficient, approximate cut-finding algorithms that balance accuracy with runtime, while still accounting for hardware constraints and sampling overhead.

\noindent\textbf{Insight 2: Exploiting Quantum Circuit Modularity ---} Quantum algorithms are typically implemented through quantum circuits whose size scales with the complexity of the target problem. These circuits often exhibit recurring patterns across qubit registers or layers. Benchmark suites QASMBench \cite{li2023qasmbench} and SupermarQ \cite{tomesh2022supermarq} show how numerous algorithm families (e.g., quantum arithmetic\cite{cuccaro2004new}, simulation\cite{suzuki2012quantum}, machine learning\cite{stein2022quclassi}, hidden subgroup\cite{bernstein1993quantum}, search\cite{grover1996fast}, optimization\cite{Kandala_2017} and variational methods\cite{tilly2022variational}) reuse the same structural blocks as qubit counts increase. Figure~\ref{fig:motivation} shows three examples with repeated modules (in green).

Ideally, leveraging modular patterns keeps cutting a 100-qubit BV circuit as manageable as a 10-qubit one. In contrast, general-purpose frameworks that treat each gate independently scale poorly with circuit size.
By identifying and exploiting this modularity, e.g., structural repetitions, our approach targets only the unique substructures within the circuit, effectively amortizing the cut-finding cost. This reduces the need for exhaustive search across the entire circuit graph, mitigating the factorial growth in complexity. Instead of treating each gate instance independently, our system constructs a compressed circuit information graph and reuses cutting decisions across all instances of the same module.

\noindent\textbf{Insight 3: Limitations of Classical Graph Algorithms ---}
Classical graph algorithms, such as subgraph-isomorphism solvers or frequent subgraph mining (FSM) techniques, are not well-suited for quantum circuit cutting. First, subgraph-isomorphism approaches require predefined patterns, which are not available in the circuit cutting context. Additionally, conventional graph properties, like node count or topology, do not directly map to quantum hardware constraints or sampling overhead.
Second, FSM algorithms typically identify recurring subgraphs within datasets and thus require significant modifications to handle single large graphs, as in circuit cutting. Even with adaptations, FSM methods such as gSpan\cite{gspan} or MoFa\cite{borgelt2002mining} exhibit exponential complexity due to the combinatorial explosion of candidate subgraphs, making them impractical for circuits beyond 100--200 gates. 

Consequently, these classical methods fail to effectively align with the practical constraints and cost models necessary for scalable and efficient quantum circuit partitioning.

\section{\sol Solution Design}
 \begin{figure*}[!htbp]
    \centering
    \includegraphics[width=0.98\linewidth]{Figures/Figures_Updated_Version/CiFold_System_Update.pdf}
    \caption{\sol Framework Overview}
    \label{fig:system}
\end{figure*} 
 
To address the challenges and limitations, we propose \sol, a scalable and graph-based circuit cutting and folding framework. In this section, we formulate circuit cutting as a constrained graph partitioning problem, introducing a novel weighted graph representation that unifies gate and wire cuts while quantifying sampling overhead as edge weights. \sol hinges on two pillars: (1) circuit folding, a systematic compression technique that identifies recurrent gate sequences across entangled qubits to construct a compact meta-graph; and (2) adaptive partitioning, which leverages the meta-graph’s hierarchical structure to guide minimal-overhead cut identification without exhaustive searches. Central to our approach are the folding factor and folding variance metrics, which rigorously assess compression efficiency and structural balance, enabling automated optimization of the folding process. From a system design perspective, we detail a parallelized workflow that combines dynamic program analysis for pattern detection (via Longest Consecutive Common Subsequence) with Weisfeiler-Lehman graph hashing for fast structural equivalence checks, ensuring scalability to large-scale circuits.

\subsection{Problem Formulation}
\label{subsec:graph_rep}

We represent a quantum circuit as a directed weighted graph \( G = (V, E, w) \), where:
\begin{itemize}
    \item \( V \) is the set of nodes representing qubit operands.
    \item \( E \) is the set of edges, each \( (u, v) \in E \) corresponding to a two-qubit gate or a wire (sequential gate on same qubit). The directed edges enforce the sequential gate ordering on each qubit, and bidirectional edges represent two-qubit interactions.
    \item \( w: E \to \mathbb{R}^+ \) assigns sampling overhead as edge weights.
\end{itemize}

\noindent
\textbf{Cutting Overhead:} Wire cuts have a fixed weight of 16. For gate cuts, the overhead depends on the entangling strength. A controlled-CZ rotation with angle \(\theta\) incurs:
$ w(e) = \left(1 + 2\sin\left(\frac{\theta}{2}\right)\right)^2.$ 
Typical controlled gates (X, Y, Z) have overhead 9, while SWAP has 49.

\noindent
\textbf{Partitioning:} A valid partitioning \(\mathcal{P} = \{P_1, \dots, P_k\}\) satisfies:
$
\bigcup_{i=1}^{k} P_i = V, \quad P_i \cap P_j = \emptyset \;\; \forall i \neq j.
$ 
Each partition contains nodes \(V_i \subseteq V\) and edges \(E_i = \{ (u, v) \in E \mid u, v \in V_i \}\). The cut set is:
$ 
E_{\text{cut}} = \{ (u, v) \in E \mid u \in P_i, v \in P_j, i \neq j \}.
$
The total sampling overhead is:
\begin{equation}
\label{equ:oh}
\Gamma_{\text{total}} = \prod_{(u,v) \in E_{\text{cut}}} w(u,v).
\end{equation}

\subsection{Meta-Graph and Folding Metrics}

Given repeated substructures, we construct a meta-graph \( G' = (V', E') \) where each \( v' \in V' \) and \( e' \in E' \) aggregates multiple instances from \( G \). Each meta-node/edge is assigned a frequency: \(\mathit{freq}(v')\), \(\mathit{freq}(e')\). By construction:
$ 
\sum_{v' \in V'} \mathit{freq}(v') = |V|, 
\sum_{e' \in E'} \mathit{freq}(e') = |E|.
$

\noindent
\textbf{Folding Factor:}
\begin{equation}
\label{equ:folding_factor}
F = \frac{|V| + |E|}{|V'| + |E'|}.
\end{equation}
A higher \(F\) indicates greater structural compression.

\noindent
\textbf{Folding Variance:} Let \(\mu_V = \frac{1}{|V'|} \sum_{v'} \mathit{freq}(v')\). The variance is:
\begin{equation}
\label{equ:folding_var}
\mathrm{Var}_{\text{fold}}(V') = \frac{1}{|V|} \sum_{v' \in V'} \mathit{freq}(v') \left( \mathit{freq}(v') - \mu_V \right)^2.
\end{equation}

\noindent
A lower \(\mathrm{Var}_{\text{fold}}(V')\) implies more uniform module sizes, simplifying partitioning. Together with \(F\), this metric helps assess folding quality and avoid \textit{under-folding} (low \(F\)) or \textit{over-folding} (high variance dominated by few large modules), ensuring a compact and balanced meta-graph for efficient circuit cutting.
 
\subsection{Framework Design: Circuit Folding}
\begin{figure}
    \centering    \includegraphics[width=0.95\linewidth]{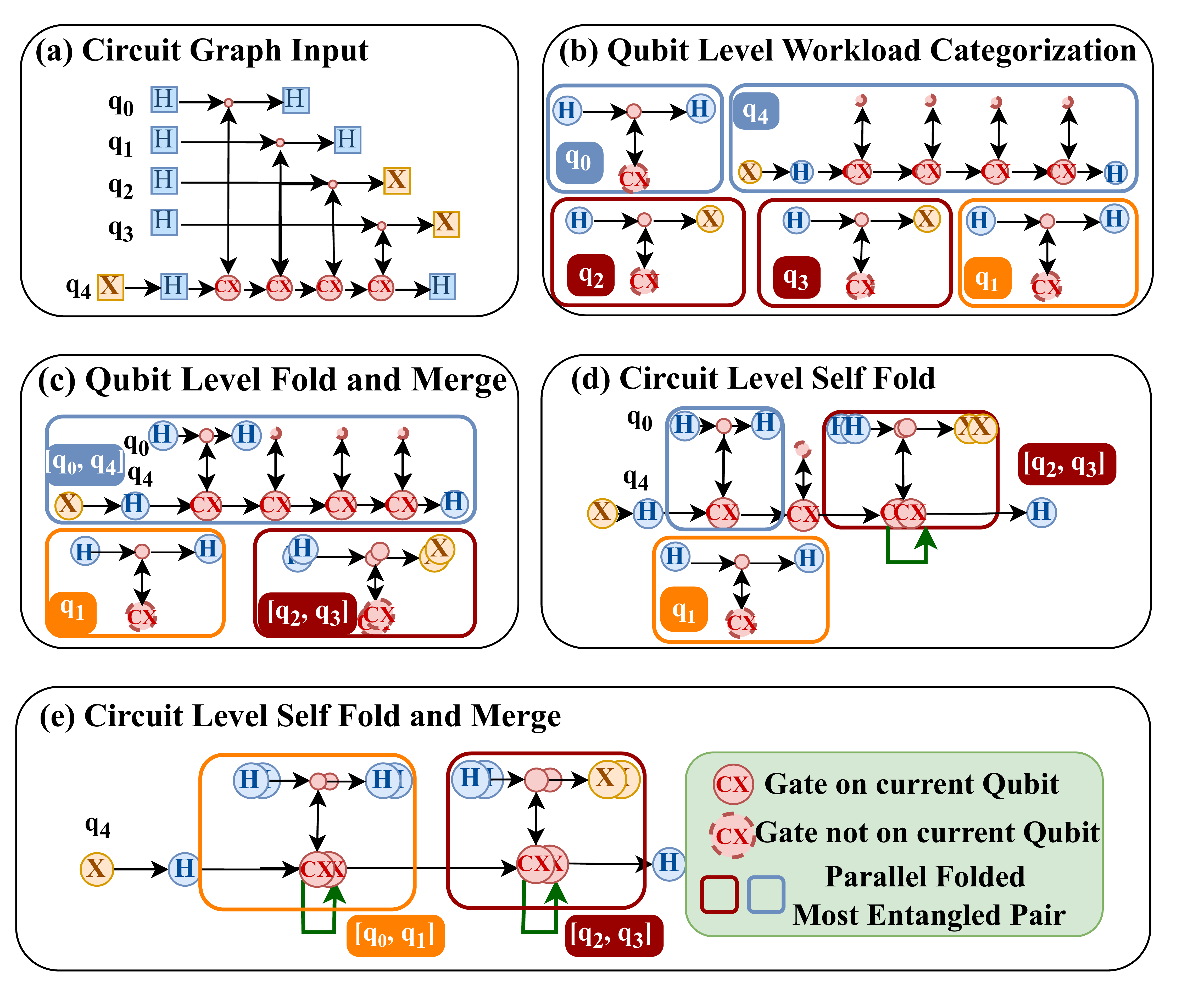}
    \caption{\textbf{Illustration of Circuit Folding:} \textbf{(a)} Circuit-level graph. \textbf{(b)} Five qubit-level gate sequences are paired based on Maximum Entangled Pairs, highlighted in matching colors, and folded in parallel. \textbf{(c)} A common gate sequence between $q_2$ and $q_3$ is detected using LCCS function, merging the red-circled nodes. \textbf{(d)} The folded structure of [$q_2, q_3$] triggers self-folding in $q_4$, marked by a green edge. \textbf{(e)} The final meta-graph reduces node count from 19 to 11.}
    \label{fig:circuit_folding}
\end{figure} 
Figure~\ref{fig:system} illustrates the \sol framework using a 5-qubit example. First, the input circuit is translated into qubit-level sequence graphs. These per-qubit graphs are then analyzed by the folding algorithm (Algorithm~\ref{alg:parallel_merge_qubits}, Figure~\ref{fig:circuit_folding}) to detect repeated modules and construct a compact meta-graph. By folding qubits with identical or similar workloads, \sol consolidates redundant structures and captures essential patterns. The resulting meta-graph is subsequently “unfolded” to generate an optimized circuit, which is finally partitioned into smaller subcircuits(Algorithm~\ref{alg:meta_graph_partition}, Figure ~\ref{fig:unfolding}). The partitioning step reduces hardware demands and sampling overhead, allowing circuits to be executed efficiently on current quantum devices.

\subsubsection{Qubit-Level Gate Sequence}

Given the graph representation of input circuit \( G = (V,E,w) \), the directed edges explicitly encode the inherent temporal ordering constraints imposed by quantum circuit operations. Leveraging these constraints, we simplify the task of detecting structural patterns by analyzing sequential gate ordering at the qubit level, thus avoiding the computational complexity associated with general subgraph isomorphism searches. Specifically, each qubit \( q_i \) yields a qubit-level sequence graph \( G_{q_i} \), as illustrated in Figure~\ref{fig:circuit_folding}(b). This extraction can be efficiently parallelized.

Given two qubit gate sequences, we employ dynamic programming to identify the Longest Consecutive Common Subsequence (LCCS). A dynamic programming table is initialized and systematically filled by comparing elements from both sequences, tracking the longest contiguous match. Only subsequences longer than a predefined threshold are retained, ensuring that the identified patterns are significant and suitable for use in the subsequent folding stage.

\begin{algorithm}[t]
\caption{Parallel Folding of Qubit-Level Pattern Graphs}
\label{alg:parallel_merge_qubits}
\begin{algorithmic}[1]
\REQUIRE Qubit-level sequence graphs $G_{q_0}, G_{q_1}, \ldots, G_{q_n}$
\ENSURE Folded meta-graph $G'$

\FORALL{$\textit{min\_LCS\_len} \in [\max(\text{seq\_len}) \dots 3]$ decreasing}
    \STATE $G_{\text{pattern}} \gets [G_{q_0}, G_{q_1}, \dots, G_{q_n}]$  
    \WHILE{$|G_{\text{pattern}}| > 1$}
        \STATE $\text{MEP} \gets \text{IdentifyMostEntangledPairs}(G_{\text{pattern}})$
        \FORALL{$(G_{q_i}, G_{q_j}) \in \text{MEP}$ \textbf{in parallel}}
            \STATE $[L_i, L_j] \gets \text{LCCS}(G_{q_i}.\text{nodes}, G_{q_j}.\text{nodes}, \textit{min\_LCS\_len})$
            \IF{$[L_i, L_j] \neq \emptyset$}
                \STATE $G_{\text{merged}} \gets \text{ComposeGraphs}(G_{q_i}, G_{q_j})$
                \FORALL{$(v_i, v_j) \in [L_i, L_j]$}
                    \STATE Transfer edges from $v_j$ to $v_i$
                    \STATE Fold $v_j$ into $v_i$
                \ENDFOR
                \STATE Replace $G_{q_i}, G_{q_j}$ with $G_{\text{merged}}$ in $G_{\text{pattern}}$
            \ENDIF
        \ENDFOR
    \ENDWHILE
    \STATE Compute $F$ and $\mathrm{Var}_{\text{fold}}(V')$ via Eq.~\eqref{equ:folding_factor},~\eqref{equ:folding_var}
    \IF{no improvement in $F$ and $\mathrm{Var}_{\text{fold}}(V')$}
        \STATE \textbf{return} folded graph $G'$
    \ENDIF
\ENDFOR
\RETURN Folded graph $G'$
\end{algorithmic}
\end{algorithm}

    


\subsubsection{Folding: Layered Frequent Pattern Discovery}

The folding process identifies repeated gate sequences, consolidating nodes and edges into a meta-graph. Each meta-node represents structurally equivalent groups from the original circuit, preserving gate parameters and connectivity.

Figure~\ref{fig:circuit_folding} demonstrates folding on a 5-qubit circuit. Algorithm~\ref{alg:parallel_merge_qubits} details the parallel folding method on qubit-level graphs \( [G_{q_0}, G_{q_1}, \dots]\). 
Initially, a sequence-length threshold is set (Line~1), and pairs of qubits with the most shared gates (Most Entangled Pairs, MEPs) are identified (Lines~3--4). The Longest Common Consecutive Subsequence (LCCS) algorithm detects common sequences (Line~6), enabling node collapsing (Lines~7--12). Edges are transferred to maintain structural integrity.

The folding operates iteratively, halving the number of graphs each step, thus requiring \( \log_2 n \) iterations for \( n \)-qubit circuits. Each iteration maximizes the folding factor \( F \) (Eq.~\ref{equ:folding_factor}) while minimizing frequency variance (Eq.~\ref{equ:folding_var}). The threshold is progressively lowered until no further improvement occurs.

\subsubsection{Unfolding: Modular Partitioning Search}

\begin{figure}
    \centering
    \includegraphics[width=1\linewidth]{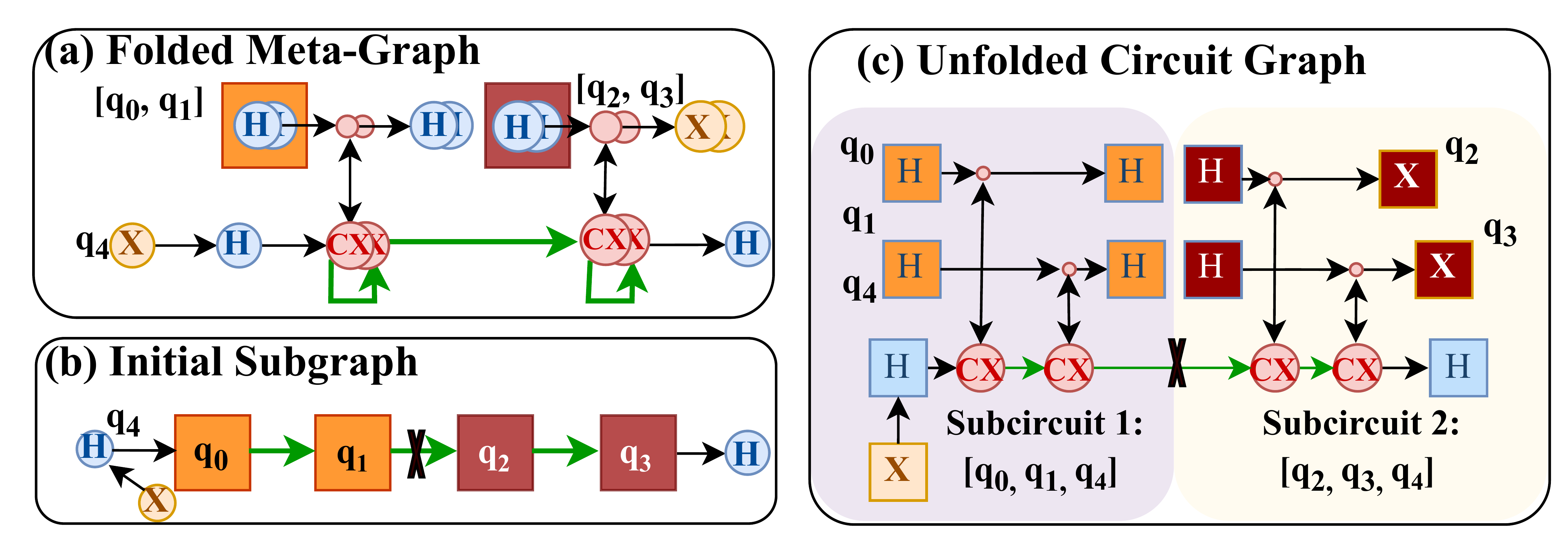}
    \caption{\textbf{Illustration of Circuit Unfolding:} The yellow and red square represents the folded node ($q_0$,$q_1$) and ($q_2$,$q_3$) in (a). (b) illustrates the identified initial subgraphs, partitioned into two 3-qubit subcircuits. (c) displays the unfolded circuit, highlighting the cuts determined in (b). }
    \label{fig:unfolding}
\end{figure}

The meta-graph encapsulating repeated structures and connectivity for partitioning. Algorithm~\ref{algo:edge_growth} details the initial partitioning strategy. An example is shown in Figure~\ref{fig:unfolding}.

\sol leverages Weisfeiler-Lehman (WL) hashing for efficiently detecting structurally similar subcircuits. Though WL hashing does not guarantee exact isomorphism, it sufficiently discriminates under quantum circuit constraints. Nodes are selected based on folding frequency, expanded using canonical edge ordering, and grouped using WL hashes.

Partitions are determined by selecting frequently occurring WL hashes. These partitions form supernodes for subsequent global partitioning, employing a METIS-inspired greedy merging and refinement strategy to minimize sampling overhead. Nodes near partition boundaries are exchangeable between partitions, further optimizing partition quality.

\begin{algorithm}[t]
\caption{Meta-Graph Guided Circuit Partitioning}
\label{alg:meta_graph_partition}
\begin{algorithmic}[1]
\REQUIRE Circuit graph $G$, Folded graph $G_f$, qubit constraint $q_{\text{con}}$
\ENSURE Partitioned subgraphs of the circuit graph

\STATE Initialize $VisitedNodes \gets \emptyset$
\STATE $P_{\text{initial}} \gets \text{ModuleFinding}(G, G_f)$ \COMMENT{Initial modules indexed by hash}
\STATE Construct graph $G'$ where each subgraph in $P_{\text{initial}}$ forms a single node
\STATE $G' \gets \text{GreedyMerge}(q_{\text{con}}, P_{\text{initial}})$
\STATE $P \gets \text{Refinement}(G', q_{\text{con}}, P_{\text{initial}})$
\RETURN Final partition $P$
\end{algorithmic}
\end{algorithm}

\begin{algorithm}[t]
\caption{Optimized Initial Partition via Edge Growth}
\label{algo:edge_growth}
\begin{algorithmic}[1]

\REQUIRE Circuit graph $G$, Folded meta-graph $G_f$
\ENSURE Initial set of partitions $P_{\text{init}}$

\STATE Initialize $\text{visited\_nodes} \gets \emptyset$, $P_{\text{init}} \gets \emptyset$

\WHILE{$|\text{visited\_nodes}| < |G.\text{nodes}|$}
    \STATE $v_{\text{freq}} \gets \text{SelectMostFoldedNode}(G_f, \text{visited\_nodes})$
    
    \FORALL{$v \in v_{\text{freq}}.\text{folded\_nodes}$ \textbf{in parallel}}
        \STATE Initialize $current\_subgraph \gets \{v\}$
        \STATE Identify candidate neighbors of $v$ in $G$
        
        \FORALL{neighbors $n$ connected via edge $e$ to $v$}
            \STATE Add node $n$ and edge $e$ to $current\_subgraph$
        \ENDFOR
        
        \STATE Compute $hash \gets \text{WeisfeilerLehman}(current\_subgraph)$
        \STATE Append $current\_subgraph$ to $P_{\text{init}}[hash]$
        \STATE Update $\text{visited\_nodes}$ with nodes in $current\_subgraph$
    \ENDFOR
\ENDWHILE

\RETURN $P_{\text{init}}$
\end{algorithmic}
\end{algorithm}

\vspace{-0.5em}
\section{Evaluation}
This section evaluates the effectiveness of the proposed \sol framework against state-of-the-art circuit cutting methods across multiple dimensions: execution time, number of cuts, sampling overhead, and reconstruction fidelity. 
\subsection{Meta Graph Comparison}
\begin{figure}[t]
    \centering
    \begin{subfigure}{0.24\linewidth}
        \centering
        \includegraphics[width=\linewidth]{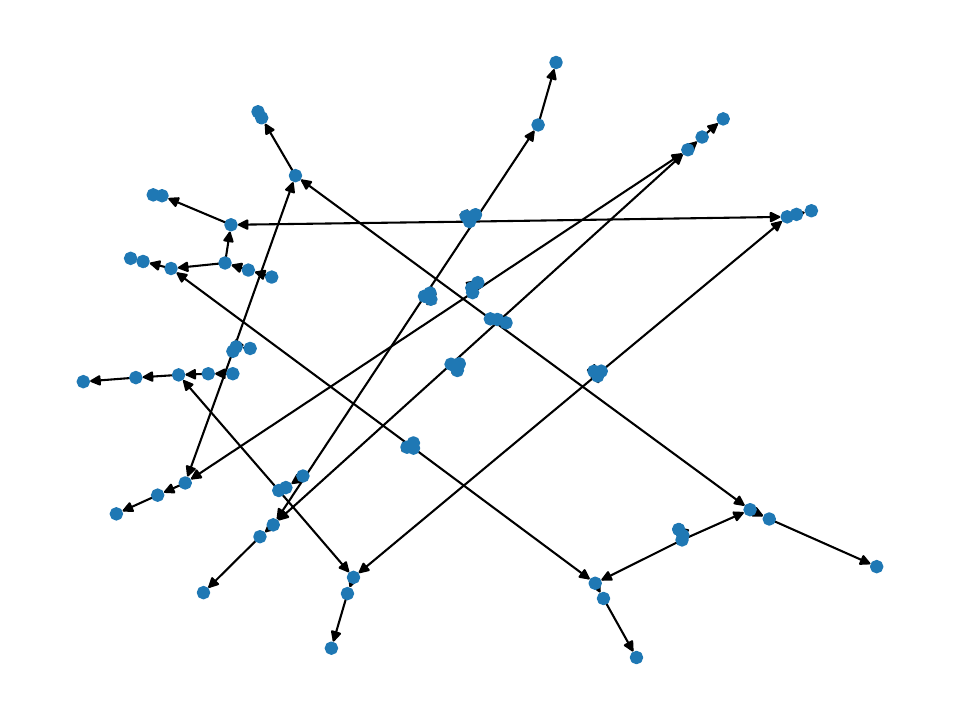}
        \caption{12-q QAOA}
    \end{subfigure}
    \begin{subfigure}{0.24\linewidth}
        \centering
        \includegraphics[width=\linewidth]{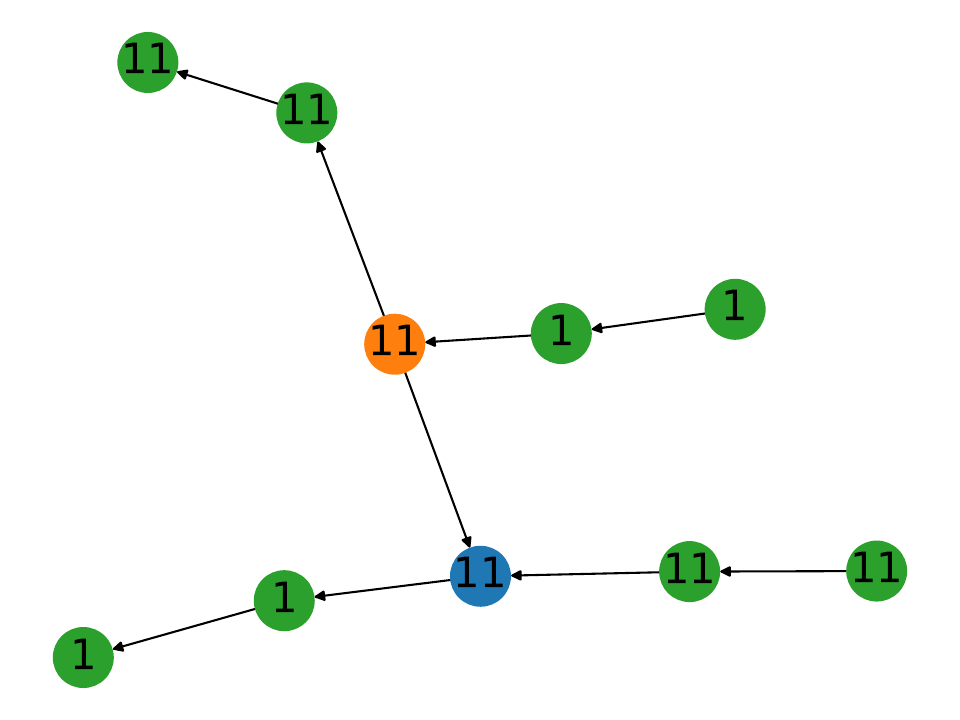}
        \caption{12-q QAOA}
    \end{subfigure}
    \begin{subfigure}{0.24\linewidth}
        \centering
        \includegraphics[width=\linewidth]{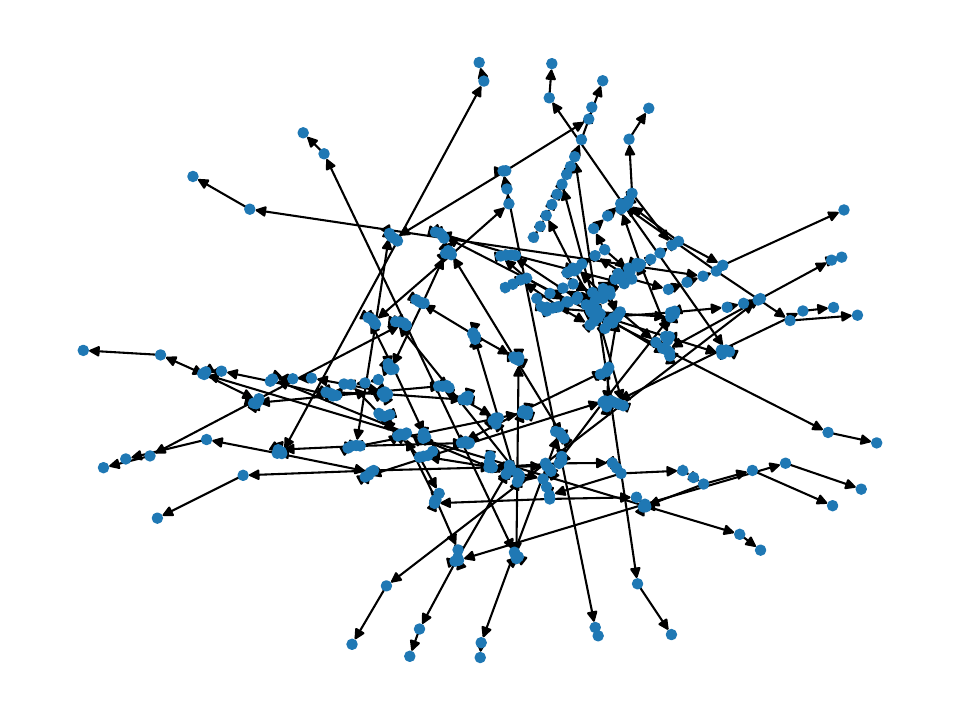}
        \caption{24-q QAOA}
    \end{subfigure}
    \begin{subfigure}{0.24\linewidth}
        \centering
        \includegraphics[width=\linewidth]{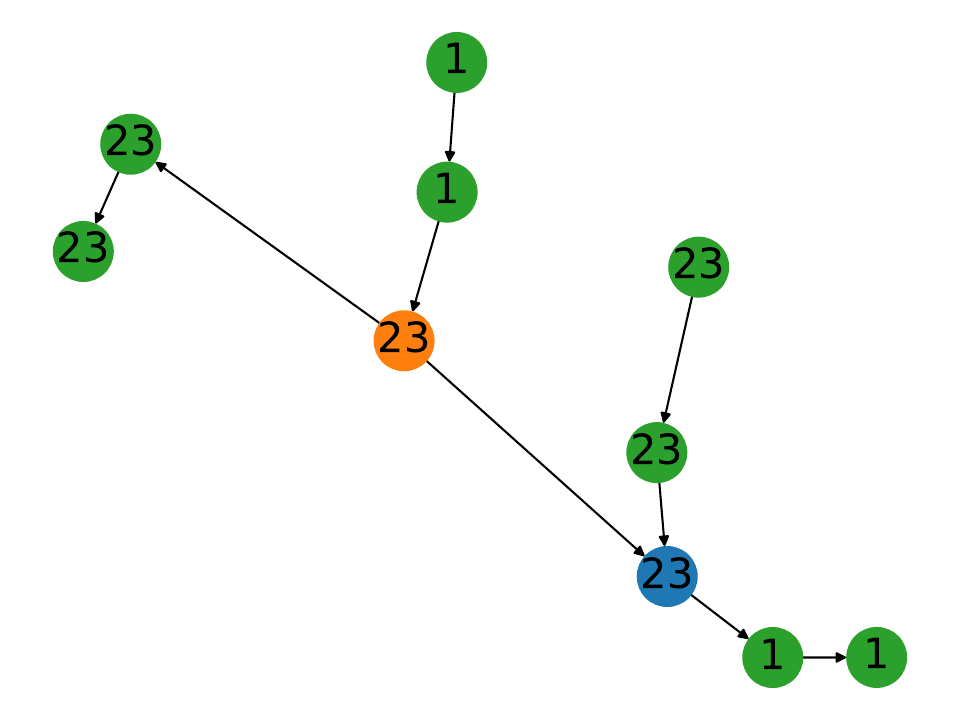}
        \caption{24-q QAOA}
    \end{subfigure}
    \begin{subfigure}{0.24\linewidth}
        \centering
        \includegraphics[width=\linewidth]{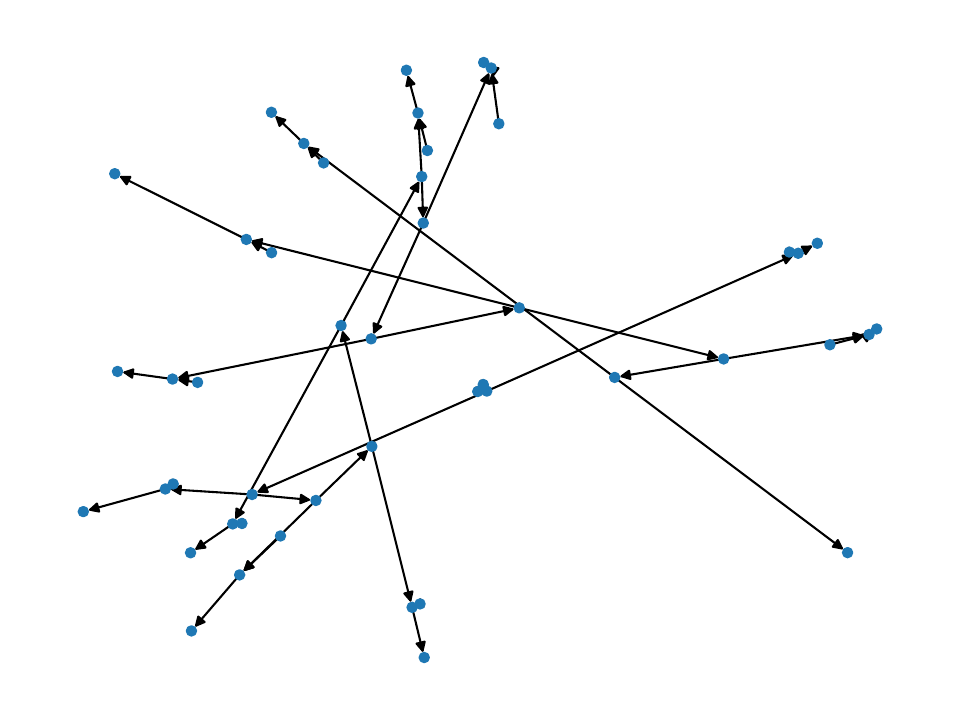}
        \caption{12-q BV}
    \end{subfigure}
    \hfill
    \begin{subfigure}{0.24\linewidth}
        \centering
        \includegraphics[width=\linewidth]{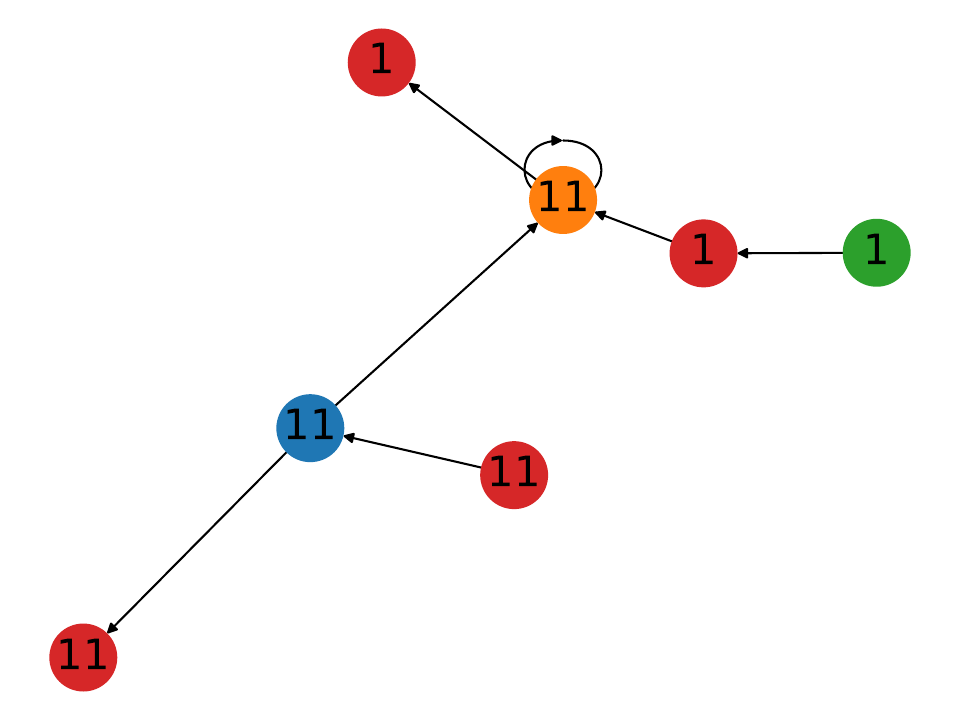}
        \caption{12-q BV}
  \end{subfigure}
    \begin{subfigure}{0.24\linewidth}
        \centering
        \includegraphics[width=\linewidth]{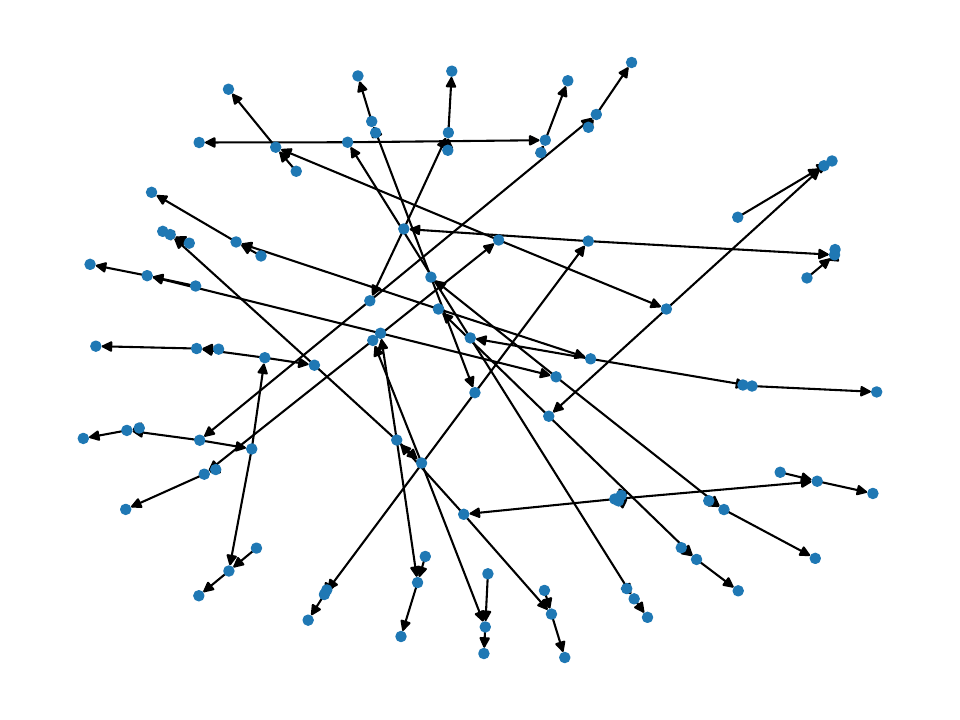}
        \caption{24-q BV}
    \end{subfigure}
    \hfill
    \begin{subfigure}{0.24\linewidth}
        \centering
        \includegraphics[width=\linewidth]{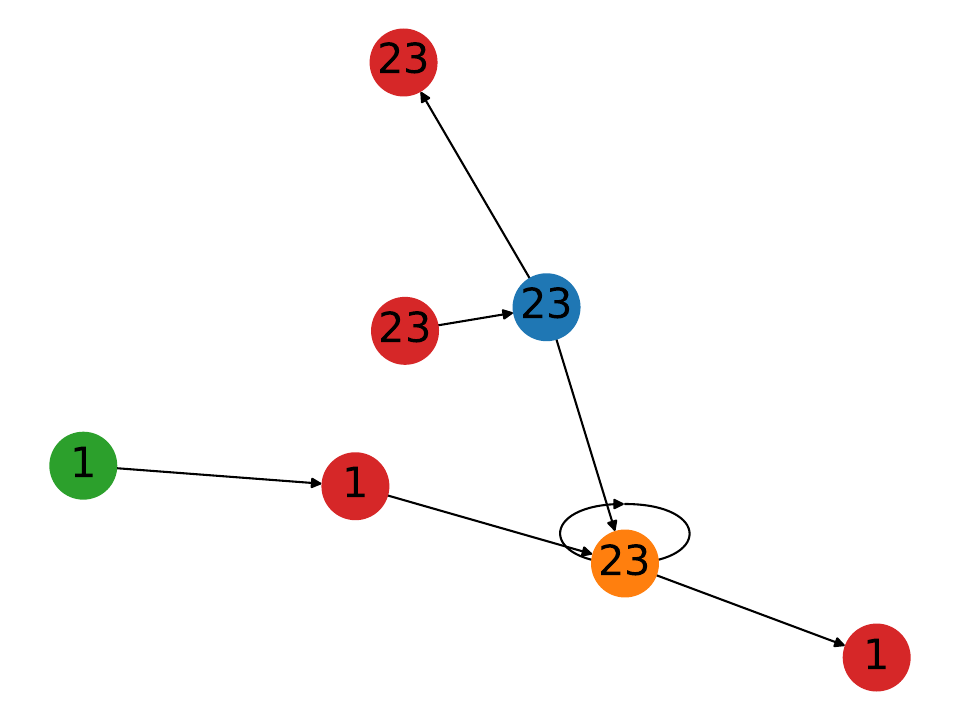}
        \caption{24-q BV}

    \end{subfigure}
    \caption{Comparison of Original Graph and Meta Graph: Each node in meta graph consolidates multiple nodes in original graph with frequency indicated. Additionally, nodes of the same color correspond to same gate types. }
    \label{fig:metagraph}
    \end{figure}

Figure~\ref{fig:metagraph} shows the graph representation of example 12 and 24-qubit circuit with its meta graph. The corresponding folding metrics in shown in \ref{tab:folding_metrics}.
These examples illustrate that a higher folding factor leads to greater reductions in graph size and complexity, while a higher folding variance indicates an uneven distribution of folding. For 12-qubit and 24-qubit circuit, they can converge to a similar meta-graph structure, differing only in frequency distributions with higher folding factor for 24-qubit circuit. Each folded node in the meta-graph aggregates multiple nodes from the original circuit, with its frequency count reflecting the extent of consolidation. Additionally, nodes of the same color correspond to identical gate types. This folding approach effectively simplifies large-scale quantum circuits while preserving their essential structure, facilitating more efficient partitioning and execution under hardware constraints.

\begin{table}[h]
    \centering
    \caption{Folding Factor and Folding Variance}
    \label{tab:folding_metrics}
    \begin{tabular}{|c|c|c|c|}
        \hline
        \textbf{Circuit Type} & \textbf{Qubit Count} & $F$  & $ \mathrm{Var}_{\text{fold}}(V')$ \\
        \hline
        QAOA & 12 & 7.32 & 24.00 \\
        BV   & 12 & 6.64 & 24.45 \\
        QAOA & 24 & 18.08 & 44.62 \\
        BV   & 24 & 13.50 & 91.08 \\
        \hline
    \end{tabular}
\end{table}

\subsection{Workload, Implementation and Settings}

\noindent \textbf{Benchmark Circuits:} We evaluate and compare \sol~using five quantum algorithms commonly employed in previous circuit-cutting studies~\cite{Tang_2021, ren2024hardware, tang2025tensorqcscalabledistributedquantum}: (1) \textbf{QAOA:} Hardware-efficient ansatz for combinatorial optimization~\cite{Kandala_2017}; (2) \textbf{BV:} Bernstein-Vazirani algorithm for exponential oracle-query speedup~\cite{bernstein1993quantum}; (3) \textbf{GHZ:} Entangled state preparation for quantum communication~\cite{greenberger1989going}; (4) \textbf{W State:} Robust multipartite entanglement; (5) \textbf{Ising:} 2-local Hamiltonian simulation for optimization~\cite{suzuki2012quantum, labuhn2016tunable}.

\noindent {\bf Implementation and Experiment Settings}: We implement the \sol with the following software: Python 3.10\cite{Python}, IBM Qiskit 1.02\cite{ibmQuantumComputing}, Qiskit-Addon-Cutting 0.9.0~\cite{qiskit-addon-cutting}, and Networkx 3.3\cite{NetworkX}. The classical components of \sol~ are executed on an AMD Ryzen 7 6800H processor running at 3.2 GHz. The quantum processors used are IBM Qiskit Emulators, using hardware calibration data from real machines including Auckland, Brisbane, CairoV2, Cusco,  WashingtonV2,  SydneyV2 and MontrealV2.

\noindent{\bf Baselines}: We compare \sol against three state-of-the-art circuit-cutting frameworks:
\noindent\textbf{Qiskit-Addon-Cutting}~\cite{qiskit-addon-cutting} utilizes a Dijkstra-based best-first search algorithm supporting both gate and wire cuts.
\noindent\textbf{CutQC}~\cite{Tang_2021} employs solver-based optimization restricted to wire cuts, guaranteeing optimal solutions for \textbf{wire-only} partitions. However, its factorial runtime growth significantly limits scalability. To manage this, a 300-second time limit is enforced, after which the best available solution (or none, if unresolved) is returned.
\noindent\textbf{FitCut}~\cite{kan2024scalable}, another \textbf{wire-only} method, combines Louvain community detection with greedy merging for efficient partitioning.


\subsection{Evaluation Metrics}

\noindent \textbf{Relative Fidelity}: We assess fidelity through an end-to-end pipeline as shown on Figure~\ref{fig:system}.  We compare the reconstructed expectation value for computational $Z$ basis from partitioned circuits against values obtained by direct, unpartitioned execution on same noisy emulators. 

\begin{figure}[!htbp]
    \centering
    \includegraphics[width=0.95\linewidth]{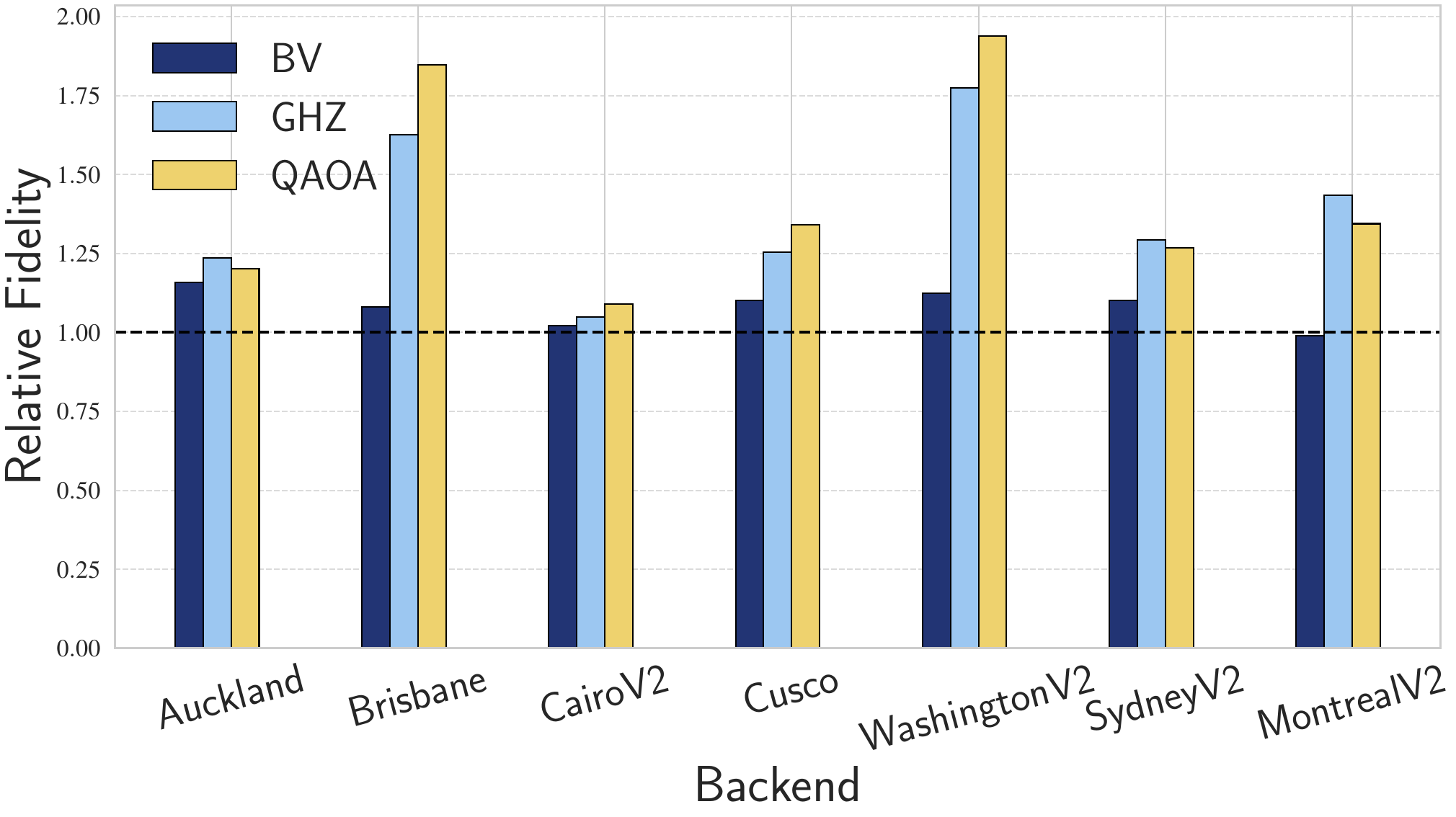}
    \caption{Relative fidelity of \sol~compared to direct execution for three 20-qubit circuits, computed as the normalized difference from uncut execution. Dashed line denotes baseline fidelity (uncut execution).}
    \label{fig:fidelity}
    \vspace{-0.1in}
\end{figure}


\noindent \textbf{Execution Time:}
Execution time reflects the overhead of partitioning, with faster runtimes indicating better scalability for large circuits. A red cross marks cases exceeding the 300-second timeout.

\noindent \textbf{Number of Cuts:}
The number of cuts directly influences the exponential classical post-processing overhead ($O(4^n)$ per wire cut). Therefore, fewer cuts indicate more efficient solutions and reduced post-processing costs.

\noindent \textbf{Sampling Overhead:}
Sampling overhead measures how many times subcircuits must be executed to achieve accurate circuit reconstruction. Hybrid approaches integrating gate and wire cuts (such as \sol~and Qiskit-Addon-Cutting) typically outperform frameworks relying exclusively on wire cuts (e.g., CutQC, FitCut).


\subsection{Performance Analysis}
\begin{figure*}[!htbp]
    \centering
    \begin{subfigure}[b]{0.32\textwidth}
        \centering
        \includegraphics[width=\textwidth]{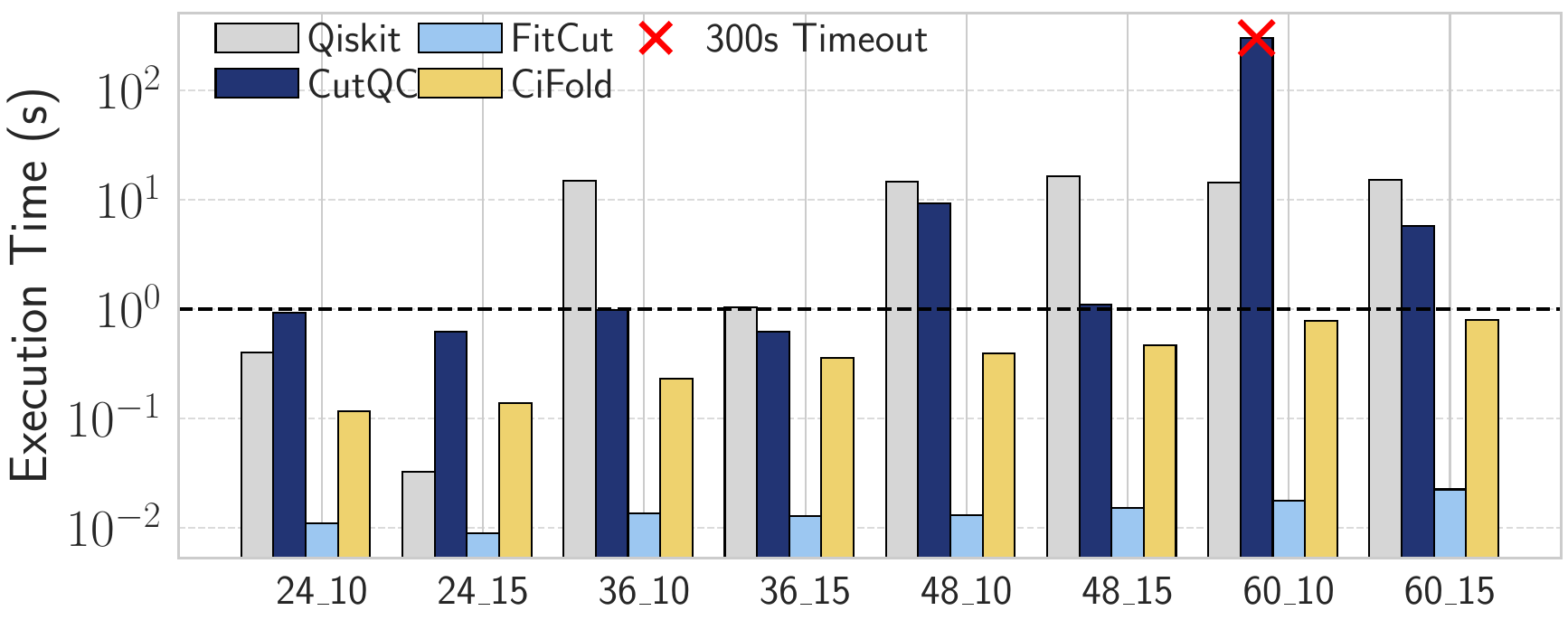}
        \caption{BV circuit - Execution Time}
    \end{subfigure}
    \hfill
    \begin{subfigure}[b]{0.32\textwidth}
        \centering
        \includegraphics[width=\textwidth]{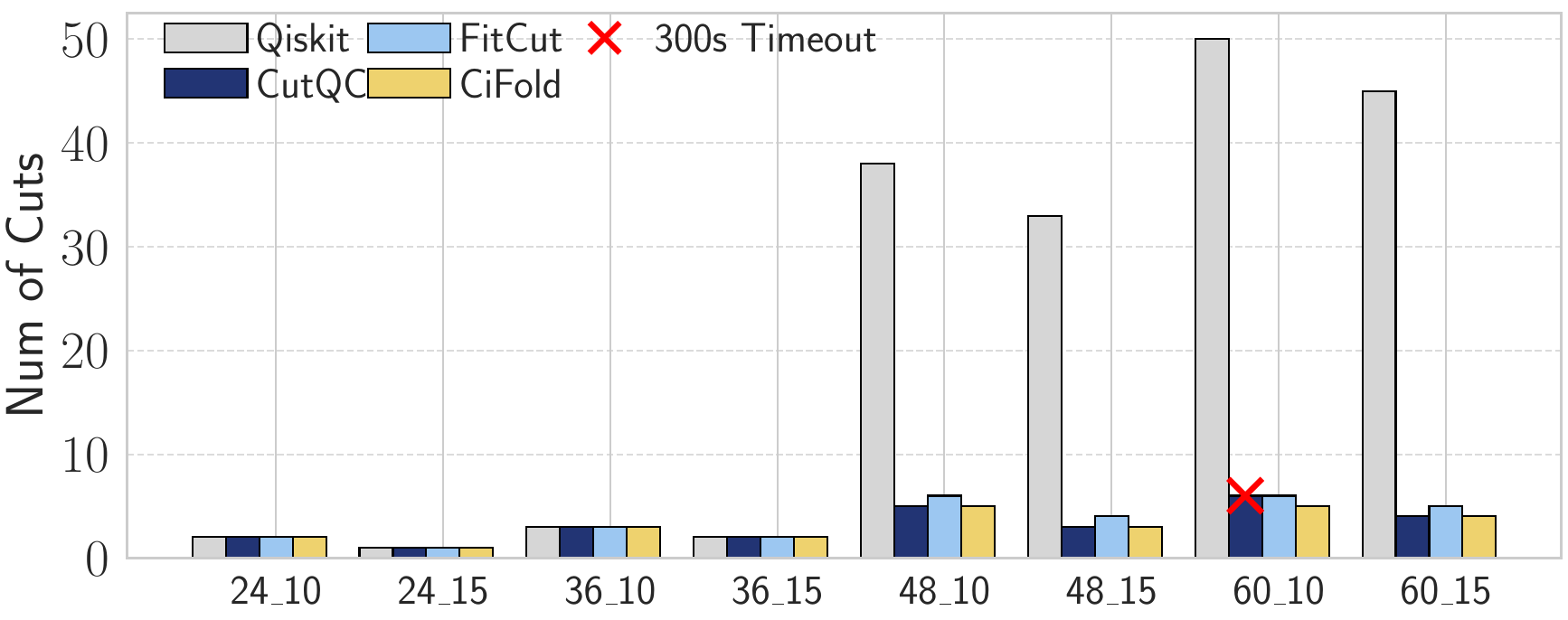}
        \caption{BV circuit - Number of Cuts}
    \end{subfigure}
    \hfill
    \begin{subfigure}[b]{0.32\textwidth}
        \centering
        \includegraphics[width=\textwidth]{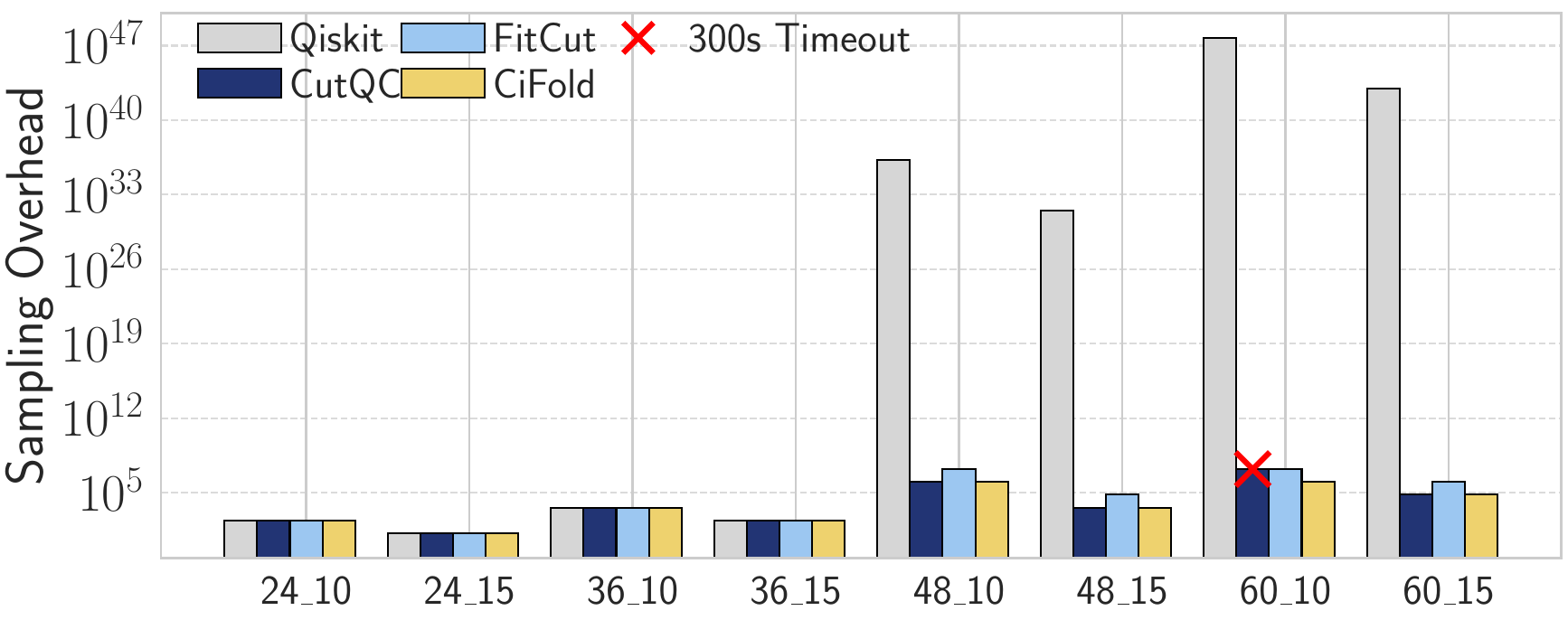}
        \caption{BV circuit - Sampling Overhead}
    \end{subfigure}

    \vspace{1em} 

    \begin{subfigure}[b]{0.32\textwidth}
        \centering
        \includegraphics[width=\textwidth]{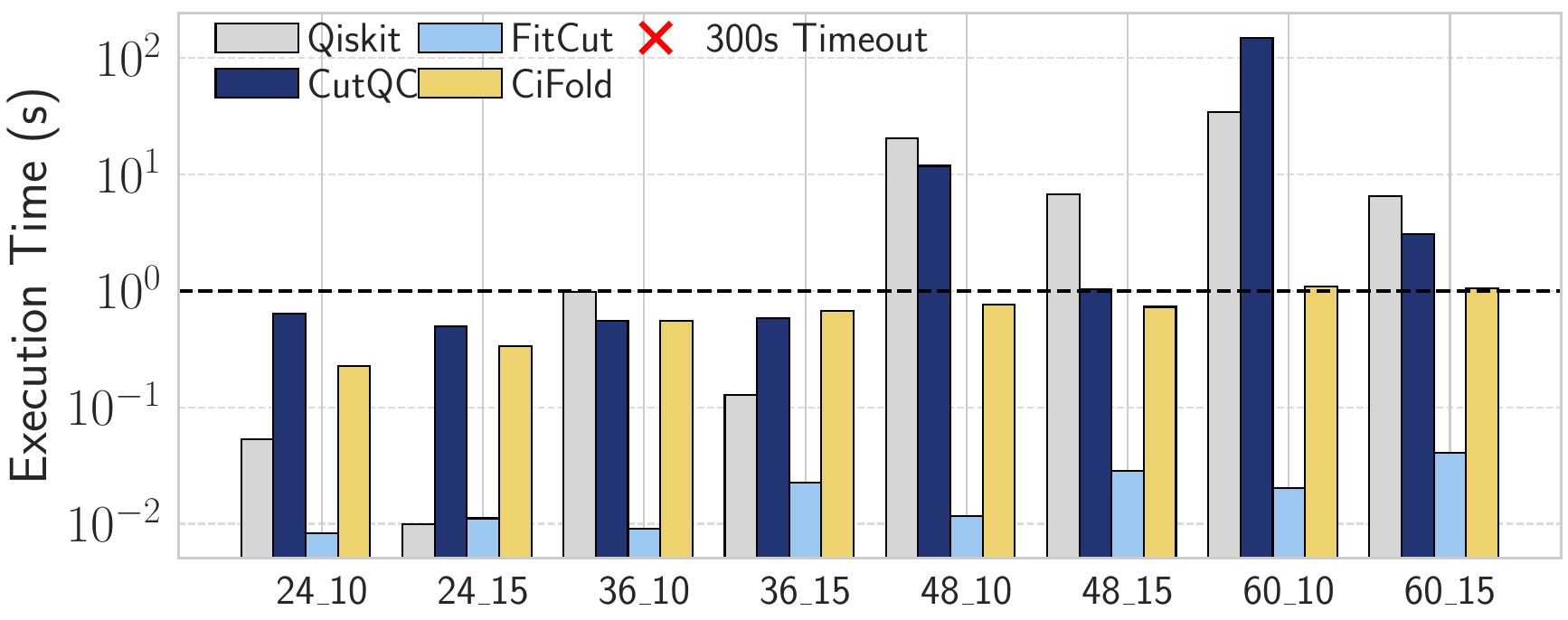}
        \caption{QAOA circuit Execution Time}
    \end{subfigure}
    \hfill
    \begin{subfigure}[b]{0.32\textwidth}
        \centering
        \includegraphics[width=\textwidth]{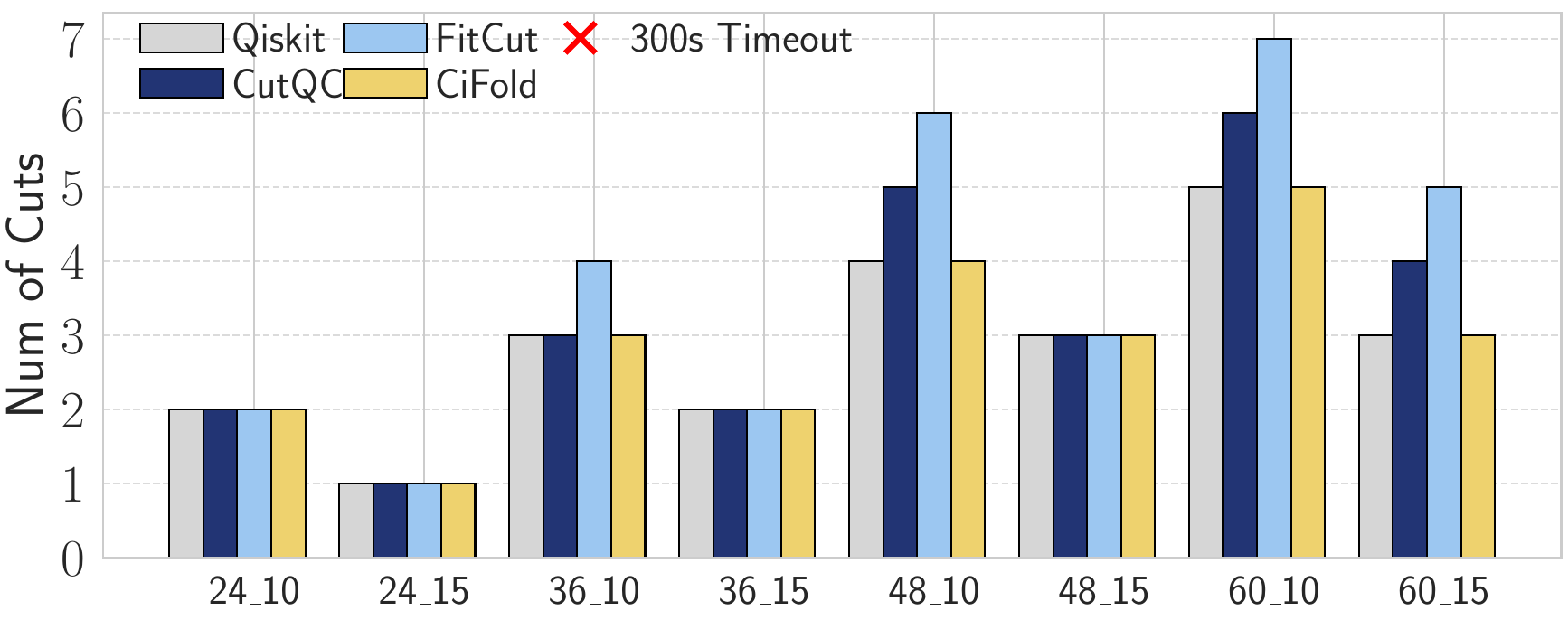}
        \caption{QAOA circuit - Number of Cuts}
    \end{subfigure}
    \hfill
    \begin{subfigure}[b]{0.32\textwidth}
        \centering
        \includegraphics[width=\textwidth]{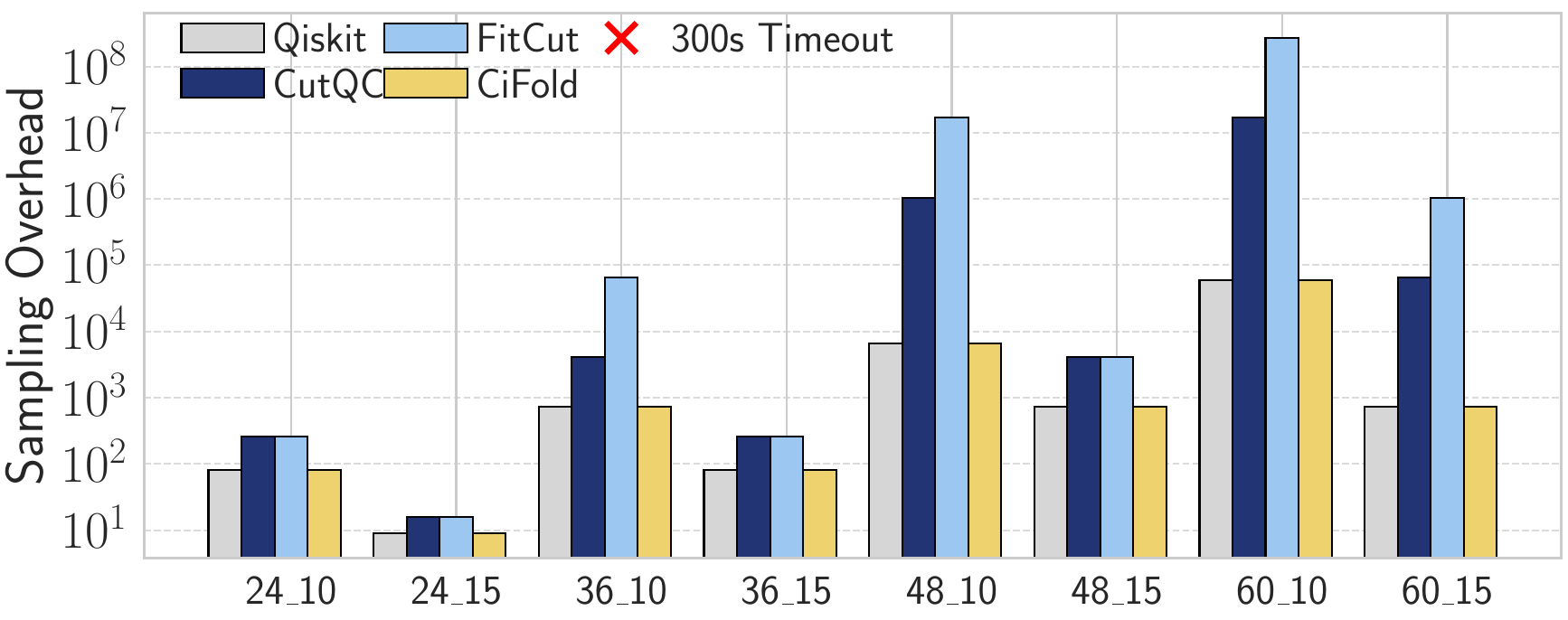}
        \caption{QAOA circuit - Sampling Overhead}
    \end{subfigure}
    \vspace{1em} 

    \begin{subfigure}[b]{0.32\textwidth}
        \centering
        \includegraphics[width=\textwidth]{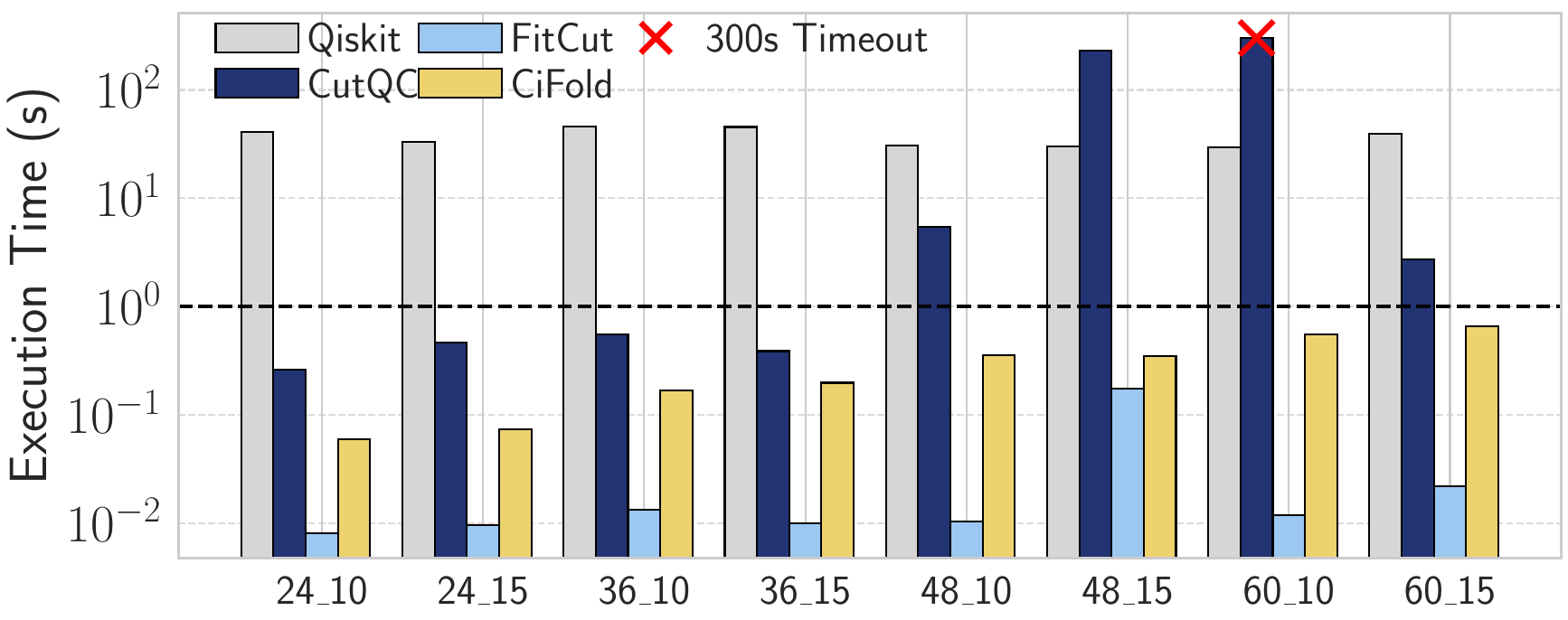}
        \caption{GHZ circuit - Execution Time}
    \end{subfigure}
    \hfill
    \begin{subfigure}[b]{0.32\textwidth}
        \centering
        \includegraphics[width=\textwidth]{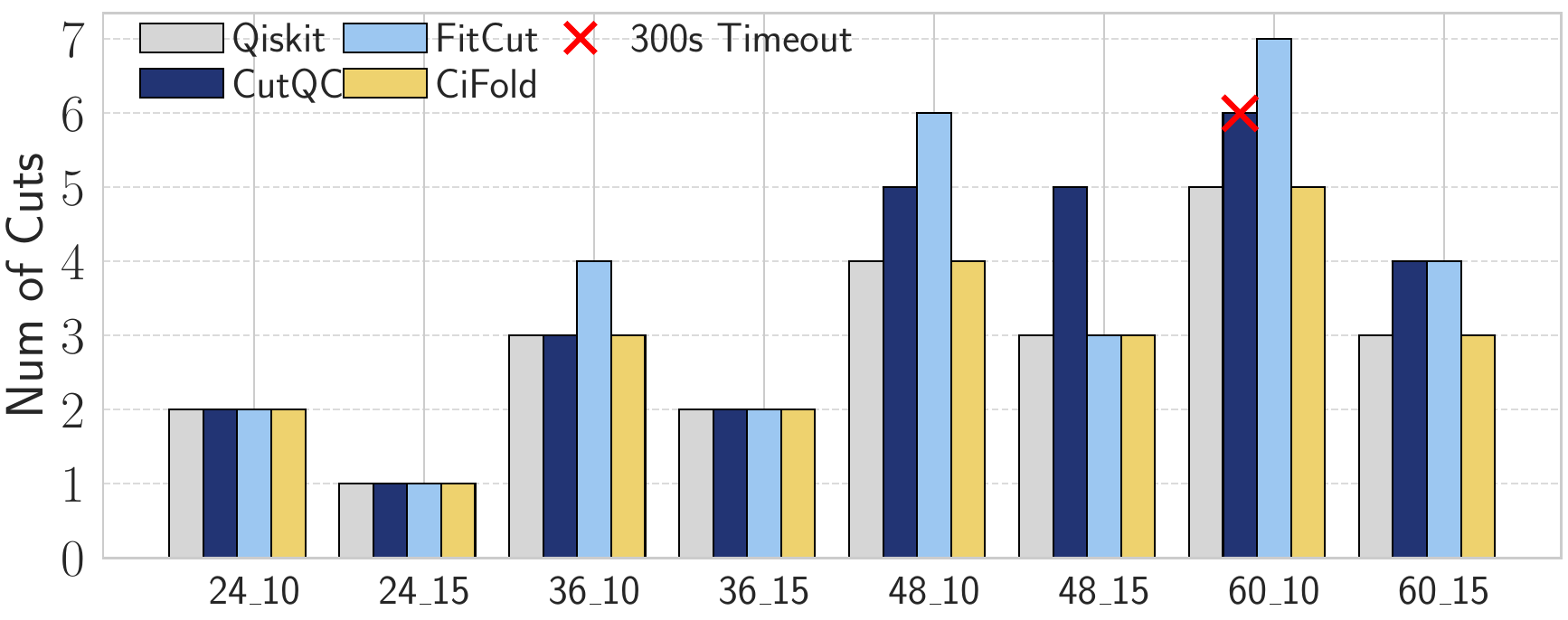}
        \caption{GHZ circuit - Number of Cuts}
    \end{subfigure}
    \hfill
    \begin{subfigure}[b]{0.32\textwidth}
        \centering
        \includegraphics[width=\textwidth]{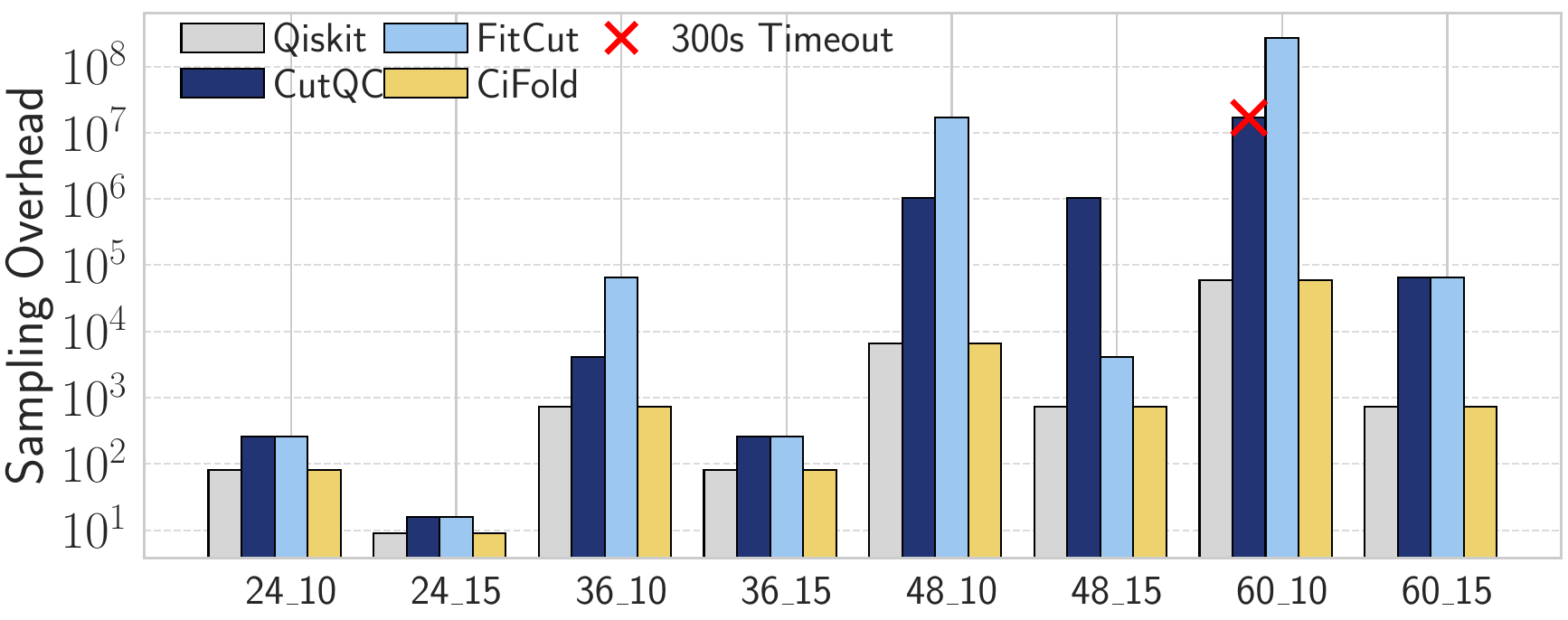}
        \caption{GHZ circuit - Sampling Overhead}
    \end{subfigure}
    \vspace{1em} 

    \begin{subfigure}[b]{0.32\textwidth}
        \centering
        \includegraphics[width=\textwidth]{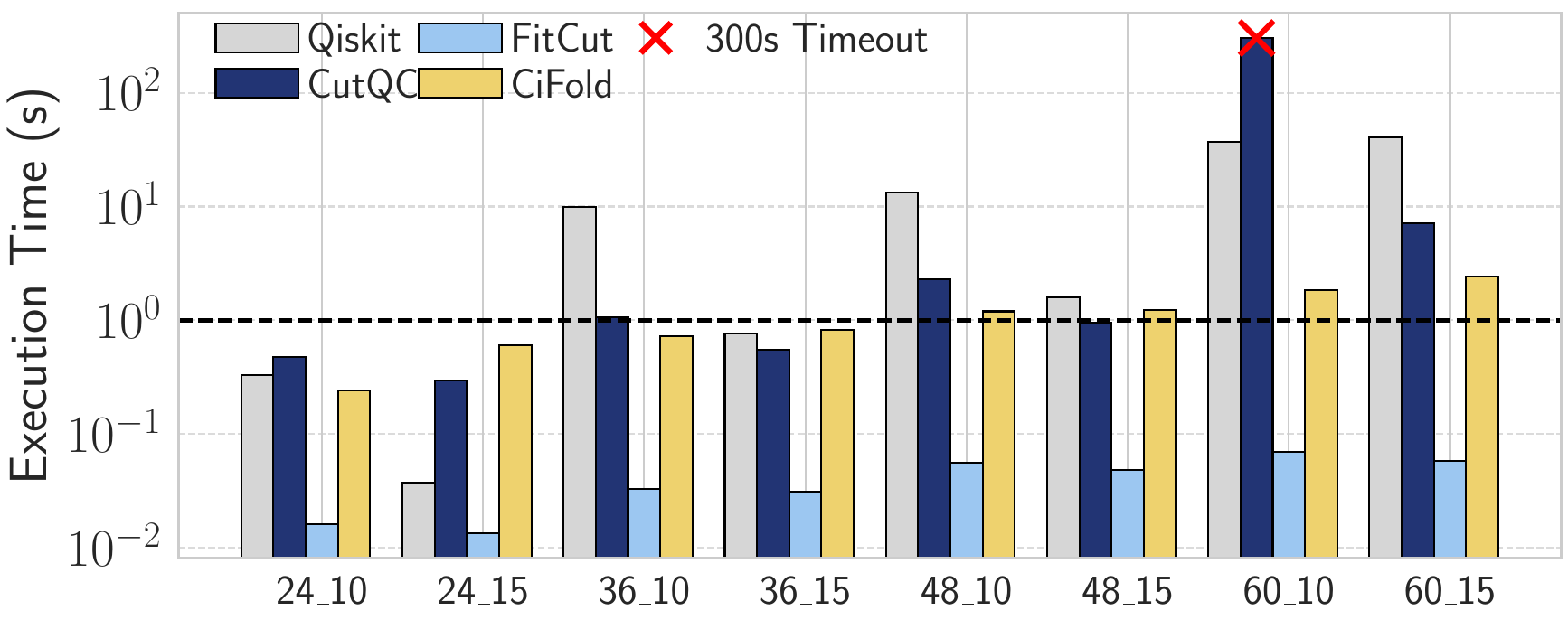}
        \caption{Ising circuit - Execution Time}
    \end{subfigure}
    \hfill
    \begin{subfigure}[b]{0.32\textwidth}
        \centering
        \includegraphics[width=\textwidth]{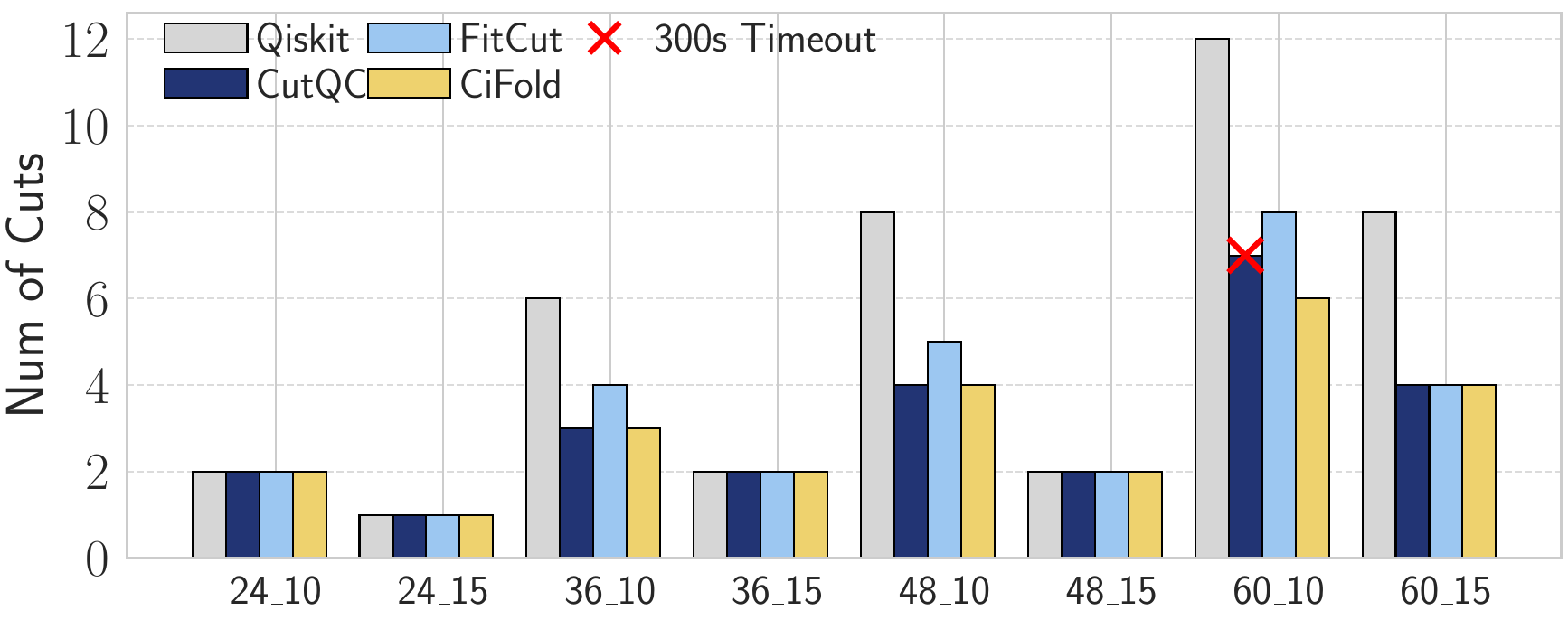}
        \caption{Ising circuit - Number of Cuts}
    \end{subfigure}
    \hfill
    \begin{subfigure}[b]{0.32\textwidth}
        \centering
        \includegraphics[width=\textwidth]{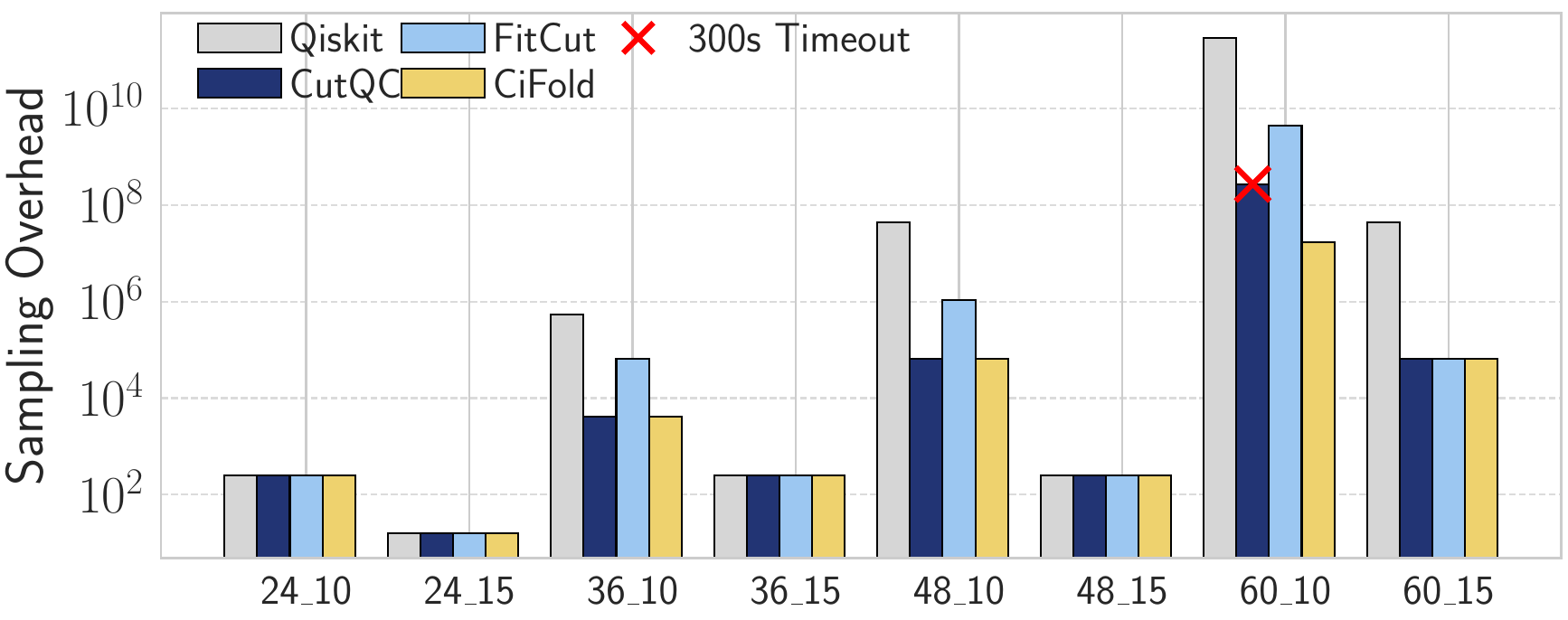}
        \caption{Ising circuit - Sampling Overhead}
    \end{subfigure}

    \vspace{1em} 

    \begin{subfigure}[b]{0.32\textwidth}
        \centering
        \includegraphics[width=\textwidth]{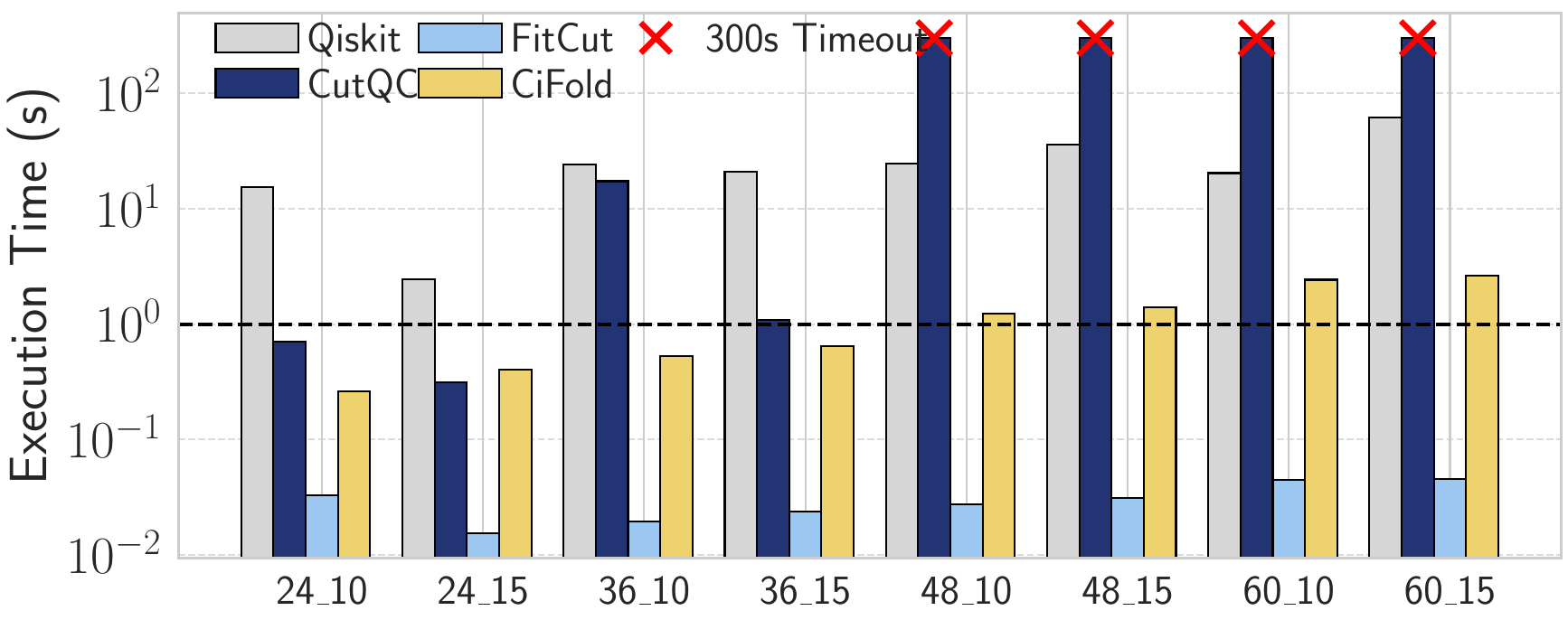}
        \caption{W State circuit - Execution Time}
    \end{subfigure}
    \hfill
    \begin{subfigure}[b]{0.32\textwidth}
        \centering
        \includegraphics[width=\textwidth]{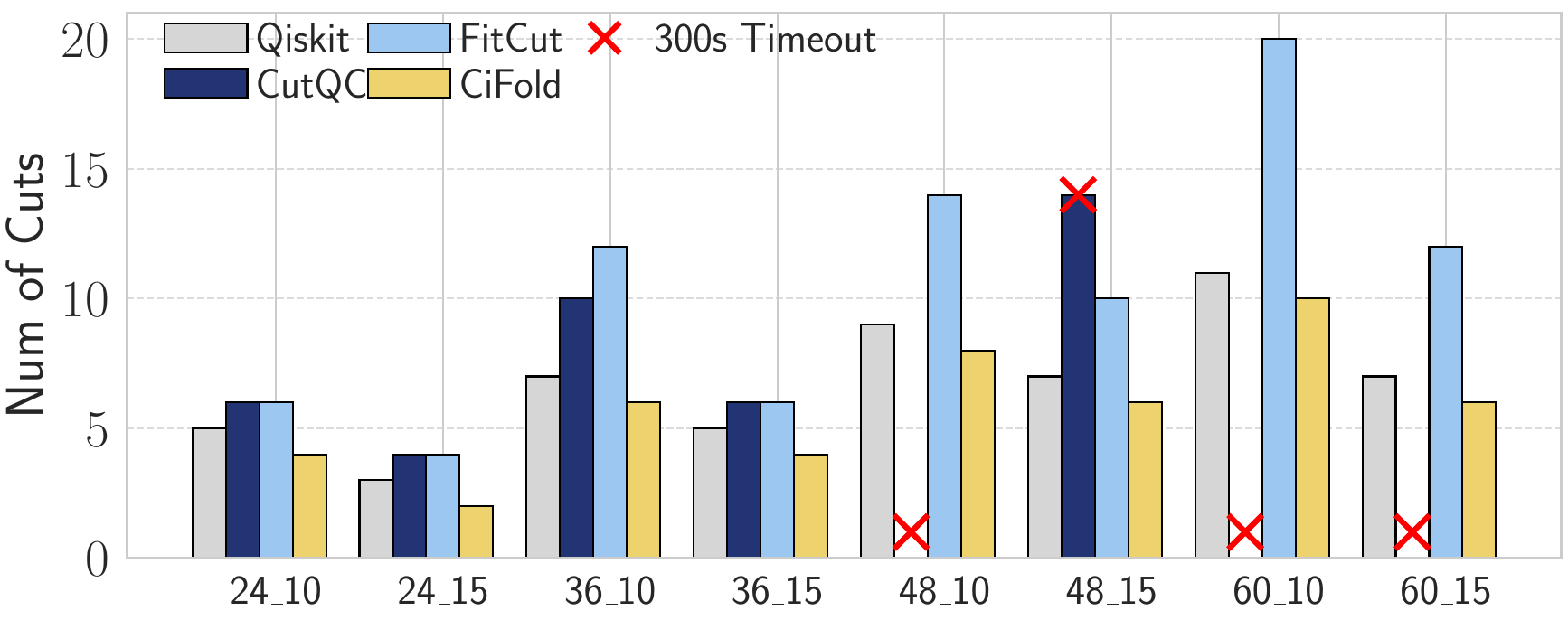}
        \caption{W State circuit - Number of Cuts}
    \end{subfigure}
    \hfill
    \begin{subfigure}[b]{0.32\textwidth}
        \centering
        \includegraphics[width=\textwidth]{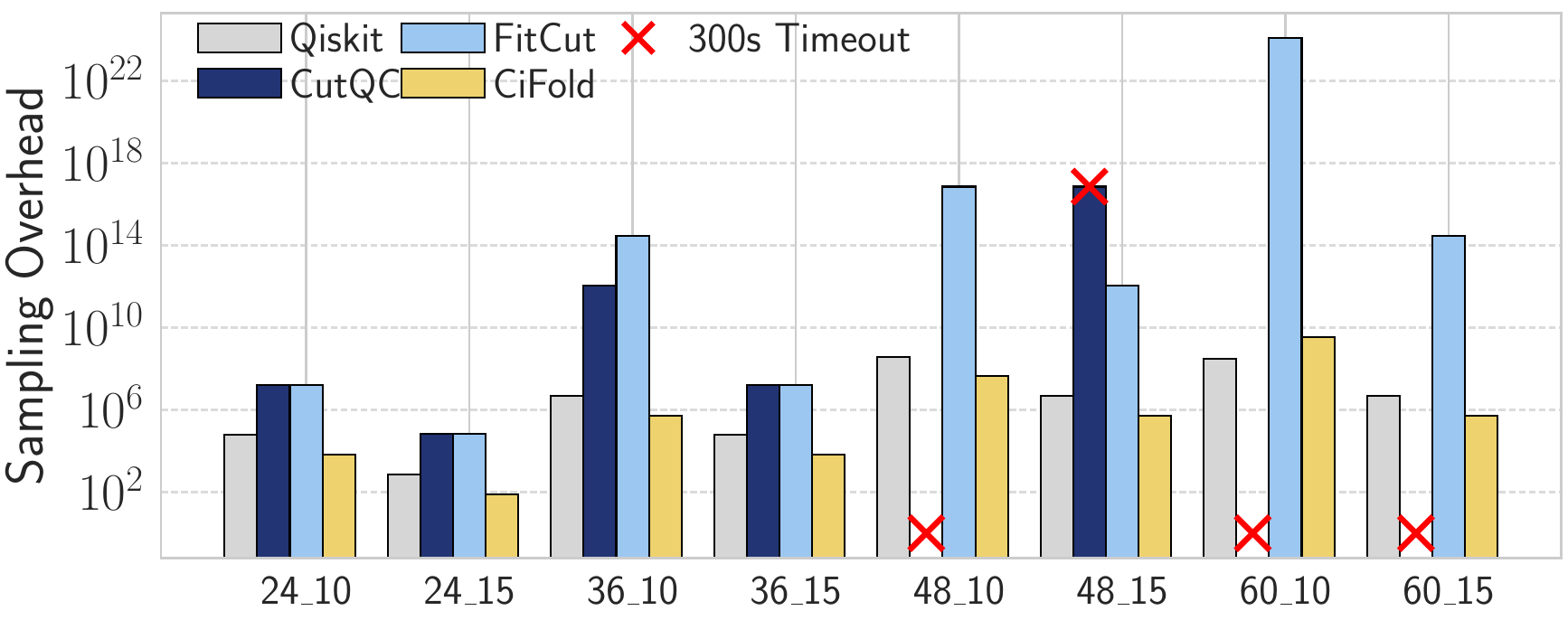}
        \caption{W State circuit - Sampling Overhead}
    \end{subfigure}
    \caption{Performance Benchmarks. The x-axis labels follow the format \{\text{size}\}\_\{\text{constraint}\}, covering circuit sizes from 24 to 66 qubits under 10-qubit and 15-qubit constraints. The dashed line represents the 1-second threshold and red cross indicates a 300 second timeout.}
    \label{fig:eval}
\end{figure*}

\noindent
\textbf{Relative Fidelity}:
In Figure~\ref{fig:fidelity}, we present the relative fidelity of reconstructed expectation values for a 20-qubit circuit with 10-qubit constraint across BV, QAOA, and GHZ benchmarks. The evaluation is conducted on seven IBM backend emulators to account for different noise profiles. The average relative fidelity improvement factors are 19.8\%, 51.8\%, 5.3\%, 23.2\%, 61.2\%, and 22.0\% for Auckland, Brisbane, CairoV2, Cusco, WashingtonV2, SydneyV2, and MontrealV2, respectively. Therefore, \sol demonstrates an improvement in fidelity due to reduced subcircuit width and depth, consistent with findings from previous studies~\cite{Tang_2021, Gentinetta2024overheadconstrained, ren2024hardware}.

\noindent
\textbf{Execution Time}:
As shown in Fig.~\ref{fig:eval} (left), the dashed line marks the 1-second threshold, and a red cross indicates a 300-second timeout. \sol consistently maintains runtimes near or below this mark across all benchmarks, leveraging parallelized folding and unfolding processes. 
FitCut consistently demonstrates the lowest runtimes (around 0.1s). This is because it is exclusively reliant on wire cuts and simplifies graph complexity compared to the hybrid gate and wire cut approaches like \sol. Without supporting gate cuts, FitCut leads to a larger number of cuts and a significant increase in sampling overhead as shown in Fig.~\ref{fig:eval}(center and right figures).

Besides, \sol outperforms Qiskit-Addon-Cutting significantly. For example, the average execution times are 1.13s vs 12.97s (Ising), 0.30s vs 36.91s (GHZ), 1.19s vs 25.76s (W-State), 0.41s vs 9.65s (BV), and 0.68s vs 8.62s (QAOA) for \sol and Qiskit-Addon-Cutting, respectively. This represents an average 94.7\% reduction. 
CutQC matches \sol performance for smaller circuits (typically under 1s), but suffers prohibitive factorial runtime growth. For example, CutQC fails to return a cutting solution for many 60-qubit circuits due to 300s API timeouts. Specifically, with 60-qubit Ising, Qiskit-Addon-Cutting and \sol return a solution in 37.3s and 1.83, but CutQC fails with 300s timeout. Notably, CutQC's runtime grows from 2.31s to 300s(timeout) for a 10-qubit constraint with 42- and 66-qubit Ising circuits.





\noindent\textbf{Number of Cuts:}
Reducing the number of cuts directly decreases classical post-processing overhead. As shown in Fig.~\ref{fig:eval} (center), \sol consistently yields fewer or comparable cuts than other methods. A red cross indicates that, within the 300-second timeout, either the best result was found or no result was returned.

For 24-qubit BV, QAOA, GHZ and Ising circuits, all evaluated methods achieve same cut counts due to the relatively low complexity. Notably, in the 24-qubit W-State benchmark, \sol and Qiskit-Addon-Cutting only require 4 and 5 cuts respectively compared to 6 cuts for CutQC and FitCut, attributable to their hybrid gate-wire cutting capability.

Performance differences become more pronounced as circuit size increases. Qiskit-Addon-Cutting lacks sufficient optimization, resulting in significantly higher cut counts—up to 50 cuts for BV and 12 cuts for Ising—compared to \sol's 5 and 6 cuts. Despite CutQC's optimality guarantee for wire-only cutting, \sol achieves fewer cuts for 66-qubit Ising benchmark with 10-qubit constraint where CutQC requires 7 cuts and \sol identifies 6. \sol shows a clear advantage in the W-State benchmark, where CutQC times out on the 60-qubit circuit and yields 14 cuts for the 48-qubit case under a 15-qubit constraint. In contrast, \sol efficiently identifies viable solutions with just 6 cuts.
These outcomes emphasize the precision of \sol in identifying cut locations, consistently minimizing the number of cuts across various circuit architectures while also significantly reducing execution time compared to solver-based method.

Overall, \sol achieves an average reduction of 31.6\% in the number of cuts (average of 3.5 cuts), compared to Qiskit-Addon-Cutting (7.8 cuts, 55.2\% reduction), CutQC (3.9 cuts, 10.0\% reduction), and FitCut (5.0 cuts, 29.5\% reduction). Specifically, \sol substantially outperforms Qiskit-Addon-Cutting due to its limited optimization, moderately improves upon CutQC despite its theoretical wire-only optimality guarantees, and consistently surpasses FitCut, whose greedy merging approach tends to become trapped in suboptimal local minima. Extending to larger circuits beyond 60 qubits, where solver-based methods such as CutQC encounter timeouts, we expect \sol to deliver even greater performance gains.

 
\noindent
\textbf{Sampling Overhead}:
Figure~\ref{fig:eval}(right figures) present the sampling overhead (computed by Eq.\ref{equ:oh}). It directly quantifies the computational cost of circuit-cutting methods, growing exponentially with each additional cut. While strongly correlated with the number of cuts, sampling overhead also depends on the type of cuts chosen, giving hybrid gate-and-wire approaches such as \sol and Qiskit-Addon-Cutting a notable advantage. For the 24‑qubit QAOA circuit, although CutQC and FitCut match \sol’s cut count, their wire‑only sampling overhead of $256$ vastly exceeds \sol’s hybrid overhead of $81$. Similarly, for the 60‑qubit QAOA benchmark, six wire cuts incur a sampling overhead of $1.68\times10^{7}$, whereas \sol’s five hybrid cuts require only $ 5.90\times10^{4}$, further highlighting the advantage of hybrid gate–wire cutting.

Overall, \sol's 31.6\% average reduction in cuts corresponds to an average sampling overhead of $8.88\times10^{7}$, compared to Qiskit’s $1.28\times10^{46}$, CutQC’s $1.94\times10^{13}$, and FitCut’s $3.00\times10^{22}$, with an overall reduction factor of $3.55\times10^{9}$(excluding CutQC Timeouts). It underscores the advantage of hybrid gate–wire circuit‐cutting strategies.

\section{Conclusion}
\label{sec:conclusion}
In this work, we proposed \sol, a hybrid gate- and wire-cutting framework that leverages modular and repetitive structures common in quantum circuits. Utilizing graph-based folding and unfolding strategies, \sol efficiently reduces circuit partitioning complexity and sampling overhead under practical hardware constraints. We introduced metrics such as the \emph{folding factor} and \emph{weighted node folding variance} to quantify and optimize the balance between circuit compression and reconstruction accuracy.

Empirical evaluations demonstrate that \sol maintains high measurement fidelity while outperforming existing circuit-cutting methods in terms of scalability, overhead, and runtime efficiency. Future work includes integrating advanced cutting protocols (e.g., parallel cuts and classical communication) and developing adaptive strategies for dynamic circuit partitioning in distributed quantum systems. Ultimately, \sol presents a promising direction for scalable and robust quantum computing within realistic hardware environments.

\section{Acknowledgment}
This research was supported in part by the National Science Foundation (NSF) under grant agreements 2301884, 2329020, 2335788, 2343535 and NVIDIA Academic Grant Program Award. 


\bibliographystyle{IEEEtran}
\bibliography{reference}
\end{document}